\documentclass[reprint,amsmath,amssymb, aps,nofootinbib]{revtex4-2}
\usepackage{graphicx}
\usepackage{tensor}
\usepackage{xcolor}
\usepackage{hyperref}
\usepackage[]{algorithm2e}
\usepackage{soul}

\begin{document}

\title{Measuring quasi-normal mode amplitudes with misaligned binary black hole ringdowns}

\author{Halston Lim} 
\affiliation{Department of Physics and MIT Kavli Institute, Massachusetts Institute of Technology, Cambridge, MA 02139}

\author{Gaurav Khanna}  
\affiliation{Department of Physics, University of Rhode Island, Kingston, RI 02881}
\affiliation{Department of Physics, University of Massachusetts, Dartmouth, MA 02747}

\author{Scott A.\ Hughes} 
\affiliation{Department of Physics and MIT Kavli Institute, Massachusetts Institute of Technology, Cambridge, MA 02139}

\date{\today}

\begin{abstract}

In recent work, we examined how different modes in the ringdown phase of a binary coalescence are excited as a function of the final plunge geometry. At least in the large mass ratio limit, we found a clean mapping between angles describing the plunge and the amplitude of different quasi-normal modes (QNMs) which constitute the ringdown. In this study, we use that mapping to construct a waveform model expressed as a sum of QNMs where the mode amplitudes and phases are determined by the source plunge parameters. We first generate a large number of calibration waveforms and interpolate between fits of each mode amplitude and phase up to $\ell \leq 8$ and $\ell - |m| \leq 4$. The density of our calibration data allows us to resolve important features such as phase transition discontinuities at large misalignments. Using our ringdown waveform model, we then perform Bayesian parameter estimation with added white Gaussian noise to demonstrate that, in principle, the mode amplitudes can be measured and used to constrain the plunge geometry. We find that inferences are substantially improved by incorporating prior information constraining mode excitation, which motivates work to understand and characterize how the QNM excitation depends on the coalescence geometry. These results are part of a broader effort to map the mode excitation from arbitrary masses and spins, which will be useful for characterizing ringdown waves in upcoming gravitational-wave measurements.

\end{abstract}

\maketitle

\section{\label{sec:intro}Introduction}

Consider a relativistic binary that inspirals due to the backreaction of gravitational-wave (GW) emission, coalescing into a single object. If that coalesced object is a black hole, then the final GW cycles emitted from the system are black hole ringdown waves, a superposition of quasinormal modes (QNMs) of the remnant black hole. Assuming general relativity, each mode has a frequency and damping time determined by the merged hole’s mass and spin. Each mode’s amplitude is determined by how it is excited, depending on the system’s mass ratio, the spins of its members, and the orbital geometry in a potentially complicated manner \cite{Kamaretsos2012,Taracchini2014SmallWaveforms,London2014,Forteza2020SpectroscopyModes,London2020,Ma2021,Lim2019,Hughes2019}.

In recent work \cite{Apte2019,Lim2019,Hughes2019}, we explored the excitation of ringdown modes in the large mass ratio limit. This limit allows us to explore the dynamics of strong-field binaries using black hole perturbation theory (BHPT): we treat binaries as the exact Kerr solution of general relativity, plus a perturbation arising from a small orbiting object. The dynamics of the perturbation to the Kerr spacetime is then governed by the Teukolsky equation \cite{Teukolsky1973}. By using a combination of frequency- and time-domain codes \cite{Sundararajan2007,Sundararajan2008,Sundararajan2010,Zenginoglu2011} to solve the Teukolsky equation, as well as a prescription for how to model the transition of the orbiting body’s evolution from adiabatic inspiral to a dynamical plunge \cite{Apte2019}, we showed that it is fairly simple to compute and characterize the ringdown modes produced by such a large mass ratio coalescence \cite{Lim2019,Hughes2019}.

Although the large mass ratio limit does not describe the binary black holes that are being measured by GW detectors today\footnote{It should be noted that it is an open question at what mass ratio perturbation theory predictions accurately describe binary black hole waveforms. Recent work \cite{Rifat2020SurrogateBinaries} finds that perturbative waveforms, when appropriately weighted and combined, agree with numerical relativity predictions surprisingly well even at mass ratios $m_1 / m_2 \simeq 0.1$ where one might expect perturbation theory to perform very poorly.}, it is a powerful tool for surveying an important portion of the parameter space describing binary black hole coalescence. By varying parameters such as the spin of the larger black hole and the angle between the black hole’s spin and orbital plane, it becomes a relatively quick and simple exercise to survey how different ringdown modes are excited, and how their excitation varies as functions of these parameters.

In Ref.~\cite{Lim2019} (which we refer to as LKAH), we undertook such a survey, examining how different ringdown modes are excited as a function of the geometry of a black hole binary’s final plunge and coalescence; Ref.~\cite{Hughes2019} provides a synopsis of this analysis. In brief, we found that fairly simple, predictable relations govern the excitation of different modes as functions of system parameters.  Given a Kerr black hole with spin parameter $a$ and perturbing mass $\mu$, the excitation of each QNM labeled by $(\ell,m,n)$ with frequency $\sigma_{\ell m n}$ is uniquely determined by three parameters: an angle $I$ which describes the inclination of the small body’s orbital plane from the black hole’s equator, an angle $\theta_{\rm fin}$ which describes the polar angle $\theta$ at which the small body plunges into the black hole’s event horizon, and the sign ${\rm sgn}(\dot{\theta}_{\rm fin})$ of the polar angular velocity in the plunge's final moments. We describe these results in more detail in Sec.~\ref{sec:previous}.

Existing models of mode excitation often include a ``standard set'' of QNMs and assume that the ringdown is dominated by the $(\ell,m) = (2,2)$ mode with subdominant contributions from the $\ell = m$ and $\ell = m + 1$ modes \cite{Kamaretsos2012,Gossan2012,Brito2018,Bhagwat2020,Bhagwat2020DetectabilityRingdown,Isi2019,Giesler2019,JimenezForteza2020,London2020,Dhani2021ImportanceWaveform}. This assumption is fairly accurate when the binary's orbital angular momentum is aligned with the remnant black hole's spin. For example, consider the numerical relativity waveform \texttt{SXS:BBH:0305} which has parameters consistent with GW150914 \cite{Boyle2019TheSimulations}. As shown in Ref.~\cite{Zertuche2021}, 96\% of the ringdown power can be described with just the $(\ell,m,n) = (2, 2,\leq2)$ modes. However, as discussed in LKAH and recently Ref.~\cite{Li2021}, the spectrum of excited modes changes significantly when the black hole's spin is not aligned with the orbital angular momentum. When the misalignment is large, the excitation of modes outside the standard set will exceed that of the $(2,2)$ mode. More generally, we find that a superposition of prograde and retrograde modes are excited, where we define ``prograde" modes as
\begin{equation}
    {\rm sgn}(\mathfrak{R}\lbrace \sigma_{\ell m n}\rbrace) =  {\rm sgn}(m),
\end{equation}
and ``retrograde" modes as
\begin{equation}
    {\rm sgn}(\mathfrak{R}\lbrace \sigma_{\ell m n}\rbrace) =  -{\rm sgn}(m)\;.
\end{equation} 
These definitions signify how the wavefronts circulate along the azimuthal direction, following Ref.~\cite{Zertuche2021}.  The notion of a ``standard set'' of late-time ringdown modes thus ceases to be a useful concept when one begins modeling black hole mergers with highly misaligned components.

Measuring the QNM content with a model which generalizes beyond a standard set of modes is challenging due to the sheer number of possible combinations. Furthermore, a given set of excited modes is not always hierarchical, and we find there are many plunge geometries for which several modes have approximately the same importance. In a detection scenario, this means that the number and variety of modes required by the data may be unclear. This is problematic in a situation where dominant modes may be missing in a model but the model is still supported --- or cannot be excluded --- by the data. Then, the measurements will be biased. Alternately, with a source model for the mode excitation, the mode amplitudes can be expressed as a function of the source parameters which drastically reduces the degrees of freedom. Incorporating this prior information in a detection scenario allows a ringdown model to generalize beyond an assumed standard set of modes while improving the measurement of the mode amplitudes.

The measurement of multiple QNMs, referred to in the literature as black hole spectroscopy, is primarily discussed in the context of the mode frequencies. For instance, a measurement of several QNM frequencies can be used to test the no-hair theorem of general relativity \cite{Israel1967EventSpace-Times,Carter1971AxisymmetricFreedom,Hawking1972BlackRelativity,Robinson1975UniquenessHole,Mazur1982,Carullo2018,Bustillo2020,Gossan2012,Bhagwat2020,Ota2020,Thrane2017,Isi2019,Meidam2014}. Such measurements can also be used to test GR by comparing their consistency with predictions of the remnant properties from full inspiral-merger-ringdown (IMR) waveform models \cite{Abbott2020c}. Generally, resolving either the (real-valued) frequency or damping time of a subdominant mode is easier than measuring its amplitude \cite{Berti2007}. Nonetheless, the QNM amplitudes carry a lot of information about a source's properties, and it is important to understand what may be learned when they can be accurately measured. Eventually, measurements of both the mode amplitudes and frequencies should be leveraged to more completely characterize waveforms than current models which only specify the frequencies. 

In this study, we make the simplifying approximation that the remnant parameters $a$ and $M$ (and thus mode frequencies) are known. Physically, this represents the limit in which priors on $a$ and $M$, as generated by inspiral-only or IMR models, are extremely tight. Recent studies demonstrate that IMR waveform models are already able to impose tight constraints \cite{Isi2019,Abbott2020,Breschi2019IMRVirgo}. Such measurements will improve with next-generation detectors like Cosmic Explorer, Einstein Telescope, and LISA which will generally be both more sensitive and open wider frequency bands than presently operating gravitational-wave detectors \cite{Amaro-Seoane2017LaserAntenna, Evans2021ACommunity, ETSteeringCommittee2020ET2020}. While understanding the correlation between the mode frequencies and amplitudes is important, we leave such an investigation to future work.

At first order in BHPT, changes to the larger black hole's parameters $a$ and $M$ are not computed. This allows us to effectively set the final remnant BH parameters, along with the starting orbital geometry, as initial conditions in our simulations. For fixed $a$ and $M$, we then generate a large number of plunge trajectories and waveforms spanning the entire range of possible merger geometries from generically oriented, precessing binaries. While this prescription is valid for large mass ratios, as the mass ratio of the system decreases towards unity, the shift in the larger black hole's parameters will no longer be negligible. Identifying exactly where first-order BHPT starts to break down is beyond the scope of this work.  We look forward to comparing with numerical relativity simulations as higher mass ratio simulations for misaligned configurations become broadly available.

With a set of waveforms for generically oriented, precessing binaries at hand, we fit the QNM amplitudes and phases using the ringdown and build a template consisting of pure QNMs. Our template includes fundamental ringdown modes up to $\ell = 8$ and $\ell \leq |m| + 4$. We then demonstrate how such a template can be used, in principle, to measure the mode amplitudes by performing a Bayesian parameter estimation on the ringdown waveform with added white Gaussian noise. 

To characterize how well our model can recover the mode amplitudes, we inject a waveform from a large mass ratio BBH with $a/M = 0.5$ and large spin-orbit misalignment $I = 137^\circ$ which excites all five $(2,m,0)$ modes. This pattern of mode excitation is fairly typical of highly misaligned mergers; it cannot be described by existing aligned- or anti-aligned-spin templates which assume that either the prograde or retrograde modes are primarily excited. When the post-peak SNR is $25$, we find that the ringdown strain data, with the aid of source-informed priors on the mode excitation, can constrain the mode amplitudes and the source plunge parameters fairly well. In contrast, when we conduct a source-agnostic analysis and implement a greedy algorithm to identify the number and variety of modes supported by the data, we find the measurements are either biased or have large credible intervals that are uninformative.

At a lower post-peak SNR of $10$, the ringdown data cannot tightly constrain individual mode amplitudes, but may still provide broad information about the plunge geometry. For instance, the detection of a significantly excited retrograde mode would indicate a plunge where the remnant spin and orbital angular momentum are anti-aligned $\vec{S} \cdot \vec{L} < 0$. However, such a measurement depends on the splitting of the azimuthal degeneracy in the mode frequencies \cite{Berti2006}. We find that if the larger black hole has insufficient spin $a \lesssim 0.5M$, the $\ell = 2$ modes cannot be resolved from each other which prevents the plunge geometry from being constrained with the mode amplitudes.

The rest of the paper is organized as follows. Sec.~\ref{sec:previous} summarizes our previous work in Ref.~\cite{Apte2019,Lim2019,Hughes2019} and how we extract mode amplitudes from the Teukolsky waveform data. Sec.~\ref{sec:analysis} contains a description of our model and parameter estimation framework. Sec.~\ref{sec:casestudy} presents a detailed example of multi-mode parameter estimation for highly inclined merger and Sec.~\ref{sec:Nmode} describes a source-agnostic analysis for the same system. Sec.~\ref{sec:retrograde} describes the impact of black hole spin on measuring mode amplitudes. We give a concluding discussion in Sec.~\ref{sec:Discussion}.

\section{\label{sec:previous}Summary of previous work}

\subsection{\label{ssec:plunge}From plunge to waveform}

\begin{figure}
    \centering
    \includegraphics[width=.4\textwidth]{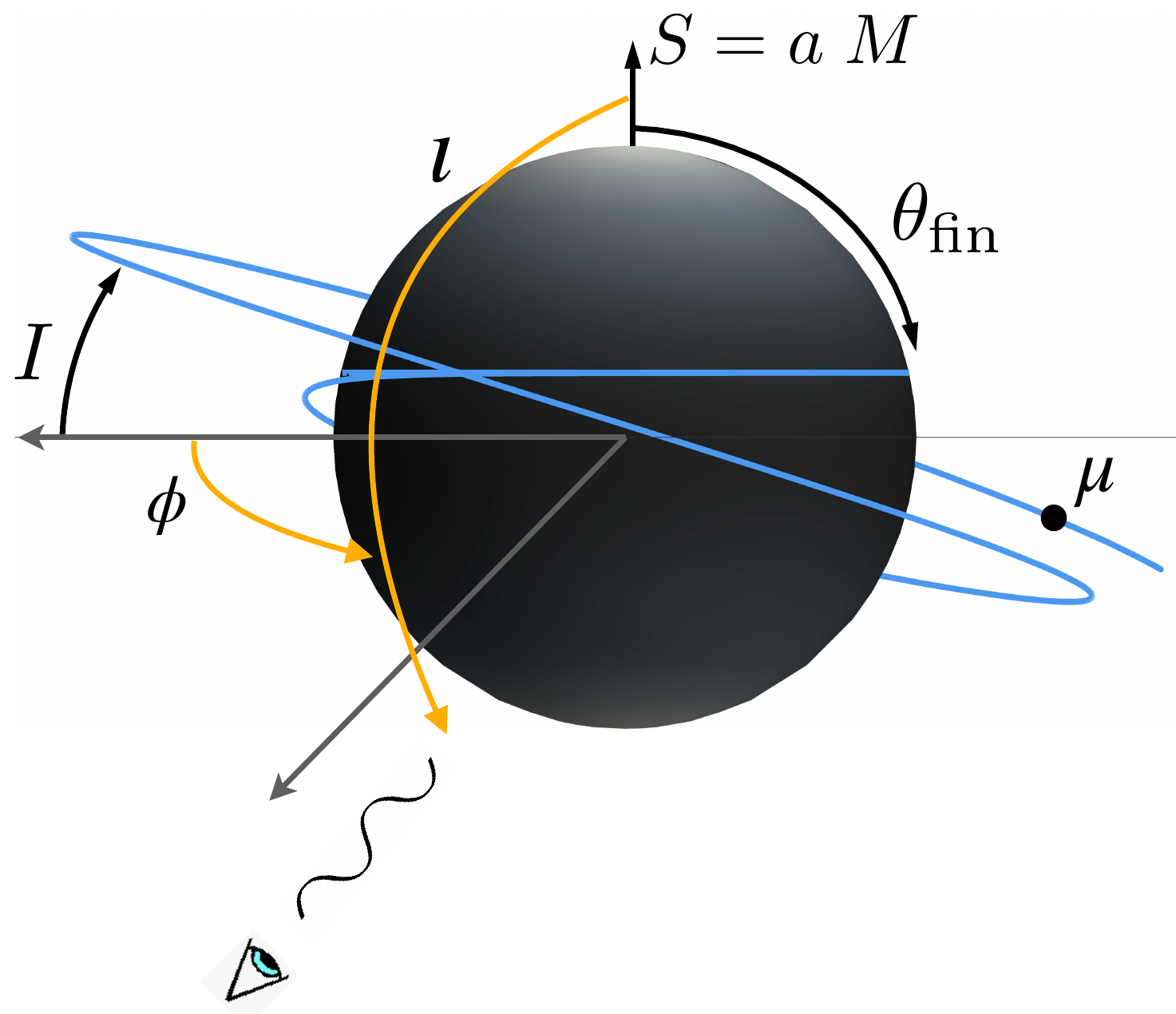}
    \caption{Parameters used to characterize the plunge geometry, $I$ and $\theta_{\rm fin}$, and angles describing the direction of gravitational radiation, $\iota$ and $\phi$. Face-on emission corresponds to $\iota = 0$; edge-on emission corresponds to $\iota = \pi/2$. We show a case where $\dot{\theta}_{\rm fin} < 0$.}
    \label{fig:diagram} 
\end{figure}

In previous work \cite{Apte2019,Lim2019,Hughes2019} we found that there is a clean mapping between the plunge geometry and ringdown mode amplitudes and phases in the large mass-ratio limit of binary coalescence. We briefly summarize the relevant results here. In Ref.~\cite{Apte2019} we develop a method to construct the worldline describing an initially adiabatic sequence of circular geodesics that transitions into a radial plunge. We parameterize each plunge with 4 parameters $\lbrace  a,I,\theta_{\rm fin},{\rm sgn}(\dot{\theta}_{\rm fin}) \rbrace$, as shown in Fig.~\ref{fig:diagram}. The spin-orbit misalignment angle $I$ is useful in characterizing the plunge geometry because it is nearly a constant \cite{Drasco2006GravitationalInspirals}, and orbit properties vary smoothly between $I = 0$ (equatorial prograde) to $I = \pi$ (equatorial retrograde). The final polar angle at which the small body freezes onto the horizon as seen by distant observers is $\theta_{\rm fin}$, and the final polar velocity during its approach is $\dot{\theta}_{\rm fin}$. The mass of the central black hole $M$ is just a scaling factor that characterizes the background spacetime which we assume is constant, while the mass of the plunging body $\mu$ is a scaling factor proportional to the amplitude of emitted gravitational waves.  Because we work in a framework that linearizes in the system mass ratio, we neglect the smaller body's spin (whose magnitude scales with the small body's mass squared) in our analysis.

Given the plunge worldline, our numerical code calculates the outgoing radiation by solving Teukolsky's equation \cite{Teukolsky1973} in the time domain and decomposes the radiation as
\begin{equation} \label{eq:sphericaldecomp}
\tensor*[]{h}{^{\rm N}}(t;\iota,\phi)=\sum\limits_{\ell ,m}\tensor*[]{h}{_{\ell m}^{\rm N}}(t)\,{_{-2}}Y_{\ell m}(\iota,\phi),
\end{equation}
where $\iota$ and $\phi$ are spherical coordinates aligned with the central black hole's spin describing the direction of the radiation (see Fig.~\ref{fig:diagram}). The superscript N emphasizes that each component is the output from our numerical code, the details of which can be found in Refs.~\cite{Sundararajan2007,Sundararajan2008,Sundararajan2010,Zenginoglu2011,Field2021AComputations}. Since our previous analysis in LKAH, several improvements have been made to the Teukolsky solver~\cite{Field2021AComputations}. First, the solver now uses a high-order, finite-difference WENO (3,5) scheme with Shu-Osher (3,3) explicit time-stepping. This scheme improves the computational efficiency of the time-domain solvers allowing for more accurate results at lower cost~\cite{Field2021AComputations}. Second, the code now uses a smoother and wider representation of the delta function. As in several previous works, we use a narrow Gaussian distribution (primarily for ease of implementation, even though it is not as efficient as a discrete delta representation). As mentioned in LKAH, for higher inclinations the worldline typically involves rather rapid movement of the smaller body over many grid points, resulting in loss of numerical accuracy. A smoother and wider representation of the delta function, used alongside the wider WENO stencils, mitigates this issue significantly.

We model the late time gravitational radiation as a superposition of quasi-normal modes,
\begin{align}
&h^{\rm RD} =
\nonumber  \\
& \frac{\mu}{d_L}\sum_{kmn} \bigg[  \mathcal{A}_{kmn} e^{-i[\sigma_{kmn}(t-t_0) - \varphi_{kmn}]}  {_{-2}}S^{a\sigma_{kmn}}_{kmn}(\iota,\phi)
\nonumber\\ 
& + \mathcal{A}'_{kmn} e^{-i[\sigma'_{kmn}(t-t_0)-\varphi'_{kmn}]}  {_{-2}}S^{a\sigma'_{kmn}}_{kmn}(\iota,\phi)\bigg]\;,
\label{eq:spheroidaldecomp}
\end{align}
where the indices $(k,m,n)$ label each quasi-normal mode with complex frequency $\sigma_{kmn}$, real-valued amplitude $\mathcal{A}_{k m n}$, and phase $\varphi_{k m n}$. We choose the fiducial start time $t_0$ based on practical considerations discussed in Sec.~\ref{ssec:extraction}. Throughout this Section and Sec.~\ref{ssec:extraction}, we use the index $k$ to distinguish the spheroidal index from the spherical index $\ell$. Every quasi-normal mode also has a ``mirror" mode of frequency $\sigma'_{kmn} = -\sigma^*_{kmn}$ with amplitude $\mathcal{A}'_{k m n}$ and phase $\varphi'_{k m n}$ \cite{Berti2006,Dhani2021ImportanceWaveform}. While the ``ordinary" modes have positive real frequencies,
\begin{equation}
    \mathfrak{R}(\sigma_{kmn}) > 0,
\end{equation}
the ``mirror" modes have negative real frequencies,
\begin{equation}
    \mathfrak{R}(\sigma'_{kmn}) < 0.
\end{equation}
We compute the QNM frequencies using the \texttt{qnm} package \cite{Stein2019Qnm:Coefficients}. 

The mirror modes have an angular dependence which can be written equivalently in two ways,
\begin{equation} \label{eq:Sangular}
    {_{-2}}S^{a\sigma'_{kmn}}_{kmn}(\iota,\phi) = (-1)^k {_{-2}}S^{a\sigma_{k-mn}}_{k-mn}(\pi-\iota,\phi)^*,
\end{equation}
which has led to different conventions in the literature associating the mirror modes with either $+m$ or $-m$ [corresponding to the $m$-index from the left and right hand side of Eq.~(\ref{eq:Sangular}), respectively]. We refer the reader to Ref.~\cite{Li2021} for a summary of these conventions. In this work, we choose our convention such that the ordinary and mirror modes labeled by the same $m$ share the azimuthal dependence $e^{-i m \phi}$. With this convention, a prograde equatorial plunge ($I = 0$) primarily excites the $\ell = 2$ prograde modes ($\mathcal{A}_{220}$ and $\mathcal{A}'_{2-20}$), whereas a retrograde equatorial plunge ($I = \pi$) primarily excites $\ell = 2$ retrograde modes ($\mathcal{A}_{2-20}$ and $\mathcal{A}'_{220}$). Our notation is similar to Ref.~\cite{Zertuche2021}, where the amplitudes in that analysis $\mathcal{A}^{\pm}_{k m n}$ [c.f. their Eq.~(4)] relate to the ones we use here as,
\begin{eqnarray}
    \mathcal{A}^+_{kmn} & = & \mathcal{A}_{k m n}, \\
    \mathcal{A}^-_{kmn} & = & \mathcal{A}'_{k m n}, \\
    \mathcal{A}^+_{k-mn} & = & \mathcal{A}_{k -m n}, \\
    \mathcal{A}^-_{k-mn} & = & \mathcal{A}'_{k -m n}.    
\end{eqnarray}
Note that this is different from the convention we previously used in LKAH. In that analysis, we used the $m$-index label on the right hand side of Eq.~(\ref{eq:Sangular}) leading the mirror modes to have a conjugated azimuthal dependence. The amplitudes and phases from LKAH map to the amplitudes and phases in this paper as
\begin{eqnarray}
    \mathcal{A}'_{k m n} & \rightarrow & \mathcal{A}'_{k -m n} \\
    \mathcal{\varphi}'_{k m n} & \rightarrow & \mathcal{\varphi}'_{k -m n} + k \pi.
\end{eqnarray}
 
In LKAH, we present a catalog of mode amplitudes normalized by $\mu / d_L$. At leading order in the small body's mass, the Teukolsky waveform amplitude is proportional to $\mu / d_L$, such that the two parameters are degenerate. 

\subsection{ \label{ssec:extraction} Mode extraction}

We construct a ringdown signal template of QNMs by building a map from the parameters describing the plunge $ \lbrace  a,I,\theta_{\rm fin},{\rm sgn}(\dot{\theta}_{\rm fin}) \rbrace$ to the QNM excitation. For each plunge trajectory, the radiation is decomposed into spherical modes $\lbrace h^{\rm N}_{\ell m} \rbrace$ which are fit to pairs of ordinary and mirror QNMs,
\begin{equation}\label{eq:algo}
    \lbrace \tensor*[]{h}{_{\ell m}^{\rm N}}(t) \rbrace \rightarrow  \lbrace \mathcal{A}_{kmn},\varphi_{k m n},\mathcal{A}'_{km n}, \varphi'_{k m n} \rbrace. 
\end{equation}
The validity of such a fitting procedure rests on several assumptions which are only approximately true; for instance, that the data comprises pure ringdown following a given truncation time, that the data can adequately constrain the modes, and that the data are a superposition of a finite number of QNMs with constant amplitudes. One way to understand by how much the data violate these assumptions is by observing how the fitted mode amplitudes vary depending on which segment of the data is fitted or which QNMs are included in the model \cite{London2014,Lim2019,Giesler2019,Bhagwat2020,JimenezForteza2020}. Characterizing this variance is important to establish the precision with which the mode amplitudes can be fitted and errors which result from imposing assumptions which are inexact.

The overfitting of non-linear behavior, numerical noise, or omitted QNMs will lead to modeling errors. One way to estimate these errors is to fit the mode amplitudes at successive times, or time intervals, and check if the fitted amplitudes approach a constant in the late ringdown. When the fits are inconsistent, some studies have suggested minimizing the residual (L2 norm) or mismatch of a least-squares fit \cite{London2014,Giesler2019,Bhagwat2020}. Such a minimization can be misleading due to overfitting, or ambiguous if there is no clear minimum. Rather than use one choice of fitting interval, we advocate for taking an average of many fits over a range in which the fit amplitudes stabilize about a constant. We find that this average over hyperparameter space helps reduces the errors in the fits when unmodeled contributions may be large. We demonstrate the effectiveness of this procedure with an example in Appendix.~\ref{app:errors}.

The quality of the fits also depends on how well the modes can be constrained by the data. For example, consider the measurement of the $m = 2$ prograde QNM amplitudes. Due to mode mixing, each spherical mode $h^{\rm N}_{\ell 2}$ may contain contributions from all ordinary $m = 2$ and mirror $m = -2$  modes which means that, in principle, these QNMs can be found by fitting $h^{\rm N}_{\ell 2}$. We find that fitting a single spherical mode with multiple QNMs does not constrain the $k \neq \ell$ QNMs well because the spherical-spheroidal mixing coefficients scale as $|a\omega|^{|\ell - k|}$ \cite{Berti2014MixingHoles}. The accuracy is improved in a multi-mode fit where many spherical modes, $h^{\rm N}_{22}, h^{\rm N}_{32}, h^{\rm N}_{42},...$. are used to fit the QNMs. {We compute the spherical-spheroidal mixing coefficients using the \texttt{qnm} package} \cite{Stein2019Qnm:Coefficients}. 

The algorithm described in LKAH is just one example of a multi-mode fitting procedure that averages over hyperparameter space. We estimate that the mode amplitudes are extracted with an absolute precision of around $\delta\mathcal{A} \sim 10^{-2}$, which is sufficient for our purposes. {For the waveforms we fit in our analysis, the errors in the fitted amplitudes are dominated by numerical errors in the waveforms, which increases with misalignment (see, e.g., fits in Fig.~\ref{fig:interp_22})}. There are also least squares-based algorithms that yield consistent results, which we also show in Appendix~\ref{app:errors}.

One additional consideration that we do not address here is the non-stationarity of spacetime. In this analysis we assume that spacetime is always Kerr plus a perturbation, and that the mass and spin of the larger black hole remain fixed throughout the inspiral, merger, and ringdown. In reality, following the merger, the mass and spin of the remnant will gradually stabilize to their final values as the spacetime approaches the Kerr metric. Fits to the QNMs will only be valid in the perturbative regime when the remnant parameters are sufficiently stabilized and other non-linearities are negligible.

High accuracy fits to the overtones pose more challenges than fits to the fundamental modes. Unlike the fundamental modes, fitting several spherical modes does not lead to tighter constraints on the overtone amplitudes. The short-lived nature of the overtones means they can only be constrained at early times when unmodeled contributions are larger. Recent work has featured fits to the overtones \cite{Zertuche2021,Lim2019,Bhagwat2020,Giesler2019}. However, as discussed in LKAH, we find the fits to the $n = 1$ overtones are only robust over a small fraction of the parameter space (for nearly aligned sources $I \leq 20^\circ$ with large spin $a \geq 0.9M$). For $n > 1$ overtones, we do not find consistent amplitudes. Fits to these higher order overtones vary significantly with the fitting interval and the number of overtones included. {There is evidence that at least one higher order overtone beyond $n > 1$ is important in comparable mass non-spinning BBH ringdowns (see e.g. Ref.~\cite{Baibhav2017}). When neglecting the $n > 1$ overtones,  fits to $\mathcal{A}_{221}$ in \texttt{SXS:0305} will vary on the order of 100\% (the difference between two fits, divided by their average) as the start of the fitting window moves from $t_{\rm peak}$ to $t_{\rm peak} + 20M$ (c.f. Fig.~3 in Ref.~\cite{Bhagwat2020}). This explains much of but not all of the modeling variations in $\mathcal{A}_{221}$. Even when including several $n > 1$ overtones and fitting at appropriately late times, fits to $\mathcal{A}_{221}$ can still vary by up to $0.33$ or 8\% as the number of fitted overtones changes \cite{Giesler2019}. While Ref.~\cite{Cotesta2022OnGW150914} suggests that this latter fitting instability might explain the discrepancy in $\mathcal{A}_{221} / \mathcal{A}_{220}$ between no-noise fits to the \texttt{SXS:0305} NR waveform \cite{Giesler2019} and posteriors calculated from GW150914 data \cite{Isi2019,Isi2021}, this difference can also be entirely explained by the omission of higher overtones but fitting from the peak.  There are additional challenges to measuring the overtone amplitudes in real detector noise. Recently, overtone detection claims in Ref.~\cite{Isi2019} have been characterized by Ref.~\cite{Cotesta2022OnGW150914} as noise-dominated, but analyses in Ref.~\cite{Isi2022} suggests that these claims are robust.} Regardless of their origin, these discrepancies should be resolved or standardized before using the overtone amplitudes for parameter estimation.

The choice of ringdown start time $t_0$ is important because the mode amplitudes scale exponentially with $t_0$ \cite{Dorband2006}. This time-shift symmetry in Eq.~(\ref{eq:spheroidaldecomp}),
\begin{eqnarray}\label{eq:timeshift}
    t_0 & \rightarrow & t_0 + \delta_0 \\ 
    \mathcal{A}_{kmn} & \rightarrow &  \mathcal{A}_{kmn} e^{-\delta_0 \mathcal{I}(\sigma_{kmn})} \\
    \mathcal{A}'_{kmn} & \rightarrow &  \mathcal{A}'_{kmn} e^{-\delta_0 \mathcal{I}(\sigma'_{kmn})} \\    
    \varphi _{kmn} & \rightarrow & \varphi_{kmn} + \delta_0 
    \mathcal{R}(\sigma_{kmn}) \\
    \varphi'_{kmn} & \rightarrow & \varphi'_{kmn} + \delta_0 \mathcal{R}(\sigma'_{kmn})    
\end{eqnarray} 
implies $t_0$, and therefore the mode amplitudes and phases, cannot be uniquely determined without introducing some arbitrary fiducial time. Our choice of the fiducial time $t_0$ is the retarded time at which the small body crosses the equivalent prograde equatorial light ring $t_{\rm LR}$, which can be solved for implicitly through
\begin{equation}\label{eq:lightring}
    r(t_{\rm LR}) = 2 M \left \lbrace 1+\cos\left\lbrack \frac{2}{3} \cos^{-1}\left(-\frac{\left|a\right|}{M}\right) \right \rbrack \right \rbrace,
\end{equation}
where $r$ is the radial coordinate of the small body in Boyer-Lindquist coordinates. The far more common choice for $t_0$ in the literature is associated with a morphological waveform peak of the $h^{\rm N}_{22}$ or $\psi_{22}^{4}$ mode (see, e.g., Refs. \cite{Giesler2019,Isi2019,Finch2021,Bhagwat2020,Ma2021,London2014}). This particular choice poorly generalizes to precessing systems because the time of the peak in the $(2,2)$ mode is discontinuous at certain plunge angles, at which the fitted mode amplitudes will also be discontinuous. At least for the quasi-circular, large mass ratio systems we consider here, the radial motion decouples from the angular motion in such a way that $t_0$ as defined by Eq.~(\ref{eq:lightring}) can be expressed as a continuous function over the plunge parameters,
\begin{equation}\label{eq:t0}
    t_0 = t_{\rm LR}\boldsymbol{(}a,M,I,\theta_{\rm fin},{\rm sgn}(\dot{\theta}_{\rm fin})\boldsymbol{)}.
\end{equation}

Numerical errors from our Teukolsky solver are larger for ``polar" orbits with extreme misalignments $ 80^\circ \lesssim I \lesssim 100^\circ$, which cause our fits in this region of parameter space less reliable. For the purposes of demonstrating our model, we consider a system in Sec.~\ref{sec:casestudy} with a inclination $I = 136^\circ$ such that the posterior samples do not overlap with this region of parameter space. In Sec.~\ref{sec:retrograde}, we calculate the Bayesian evidence, which requires calculating the likelihood marginalized over all parameters, and find that the inclusion or exclusion of the polar orbits from our prior does not significantly change the calculated evidence. Nevertheless, with future code improvements, we aim to improve the accuracy of these polar orbits.

\section{\label{sec:analysis}Ringdown Model and Analysis}

\subsection{\label{sec:model} Constructing the ringdown signal template}

Studies which map the mode excitation from source parameters have mostly focused on binaries without orbital plane precession \cite{Kamaretsos2012,Taracchini2014SmallWaveforms,London2014,Forteza2020SpectroscopyModes,London2020,Ma2021}. For instance, Ref.~\cite{London2020} provides fits to the mode excitation for mass ratio $q\equiv m_1/m_2 \le 15$ for aligned-spin systems and for the standard set of prograde modes with $\ell \leq 4$. This partial map was used to constrain the mode amplitudes in GW150914, assuming an aligned-spin source. Partial maps exist for other special cases, such as the superkick configuration of equal mass, non-precessing binaries explored in Ref.~\cite{Ma2021}. In that study, the quadrupole mode excitation was mapped from various spin tilt angles $\lbrace \phi_{\rm init}, \theta_{\rm init} \rbrace $, which can be related to the plunge parameters we use here in the large mass ratio limit $\lbrace I,\theta_{\rm fin}\rbrace$.

\begin{figure}
    \centering
    \includegraphics[width=.5\textwidth]{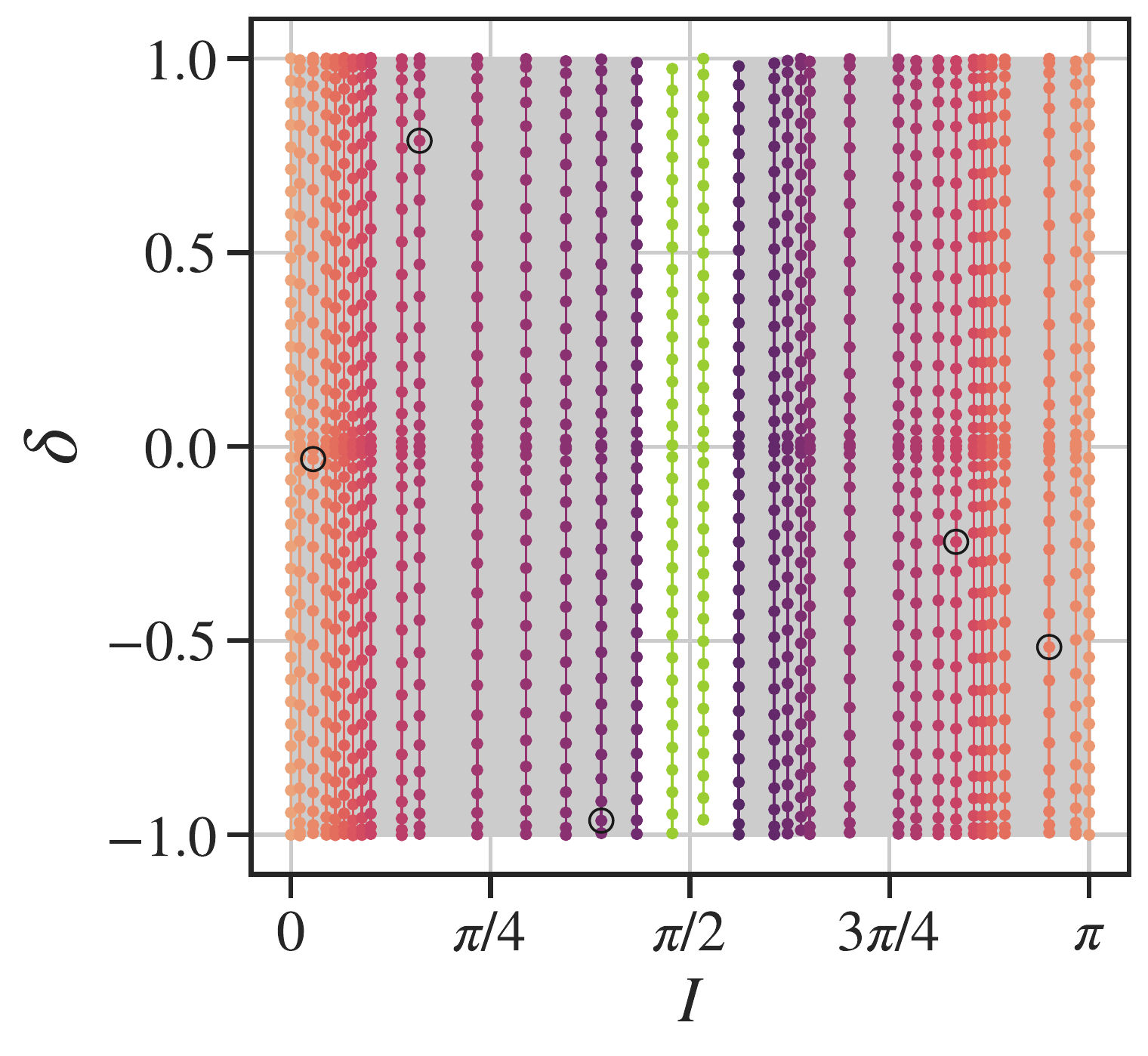}
    \caption{Plunge parameters of waveforms used to construct the ringdown template for $a/M = 0.5$. Each vertical line represents a set of waveform data with inclination $I$, consisting of 36 different $\delta$ samples. The line color indicates how misaligned the orbit is, with more misaligned orbits plotted with darker lines, and more aligned orbits are plotted with lighter lines. The left (right) shaded region is the calibration domain over which we construct our interpolating functions for prograde (retrograde) orientations. The calibration domain does not include the $I = 86^\circ$ and $93^\circ$ data (green lines), due to larger numerical errors in our waveforms. We estimate the error of our interpolation by constructing a test interpolant that excludes 5 test waveforms (open circles), and comparing the interpolated values at these test points to direct fits from those waveforms (see Fig.~\ref{fig:interp_error}). At the boundaries, $I = 0$ and $I = \pi$, the plunge parameter $\delta$ is undefined since $\theta_{\rm min} = \theta_{\rm max} = \pi/2$ [Eq.~(\ref{eq:delta})]. In other words, $\delta$, which describes at what point along the orbit the small body crosses the horizon, has no intrinsic effect on the waveform when the plunge is axis-symmetric. To include these boundaries in the interpolation, we use the fits from the equatorial prograde and retrograde plunge to populate all the points along the $I = 0$ and $I = \pi$ boundaries, respectively. We generate a similar set of simulations for the other spin cases $a/M = 0.1, 0.3, 0.7, 0.9$.}
    \label{fig:grid} 
\end{figure}

We continue this effort to map the mode excitation by considering generically aligned precessing binaries in the large mass ratio limit. Using fits from a set of calibration waveforms, we build interpolating functions for each mode amplitude and phase. We include 41 fundamental ordinary and mirror modes up to $\ell = 8$  and $\ell \leq |m|+ 4$. Instead of interpolating with the $\theta_{\rm fin}$ coordinate, we define a new parameter $\delta$ that is continuous across the $\dot{\theta}_{\rm fin} > 0$ and $\dot{\theta}_{\rm fin} < 0$ branches and normalized by the range of polar motion $\theta_{\rm max} - \theta_{\rm min} = 2I$. We define $\delta$ as
\begin{equation} \label{eq:delta}
    \delta = 
    \begin{cases} 
      \left(\theta_{\rm fin} - \theta_{\rm min}\right) / (\theta_{\rm max} - \theta_{\rm min}), & \dot{\theta}_{\rm fin} > 0 \\
      -1 + \left(\theta_{\rm max} - \theta_{\rm fin}\right) / (\theta_{\rm max} - \theta_{\rm min}), & \dot{\theta}_{\rm fin} < 0,
   \end{cases}
\end{equation}
such that $-1 \leq \delta \leq 1$. In Fig.~2, we show the plunge parameters of waveforms in our $a = 0.5M$ data set, which includes 36 $\delta$ samples (represented by points) at various $I$ values (represented by vertical lines) ranging from $I = 0$ (prograde equatorial) to $I = \pi$ (retrograde equatorial).

\begin{figure}
    \centering
    \includegraphics[width=.5\textwidth]{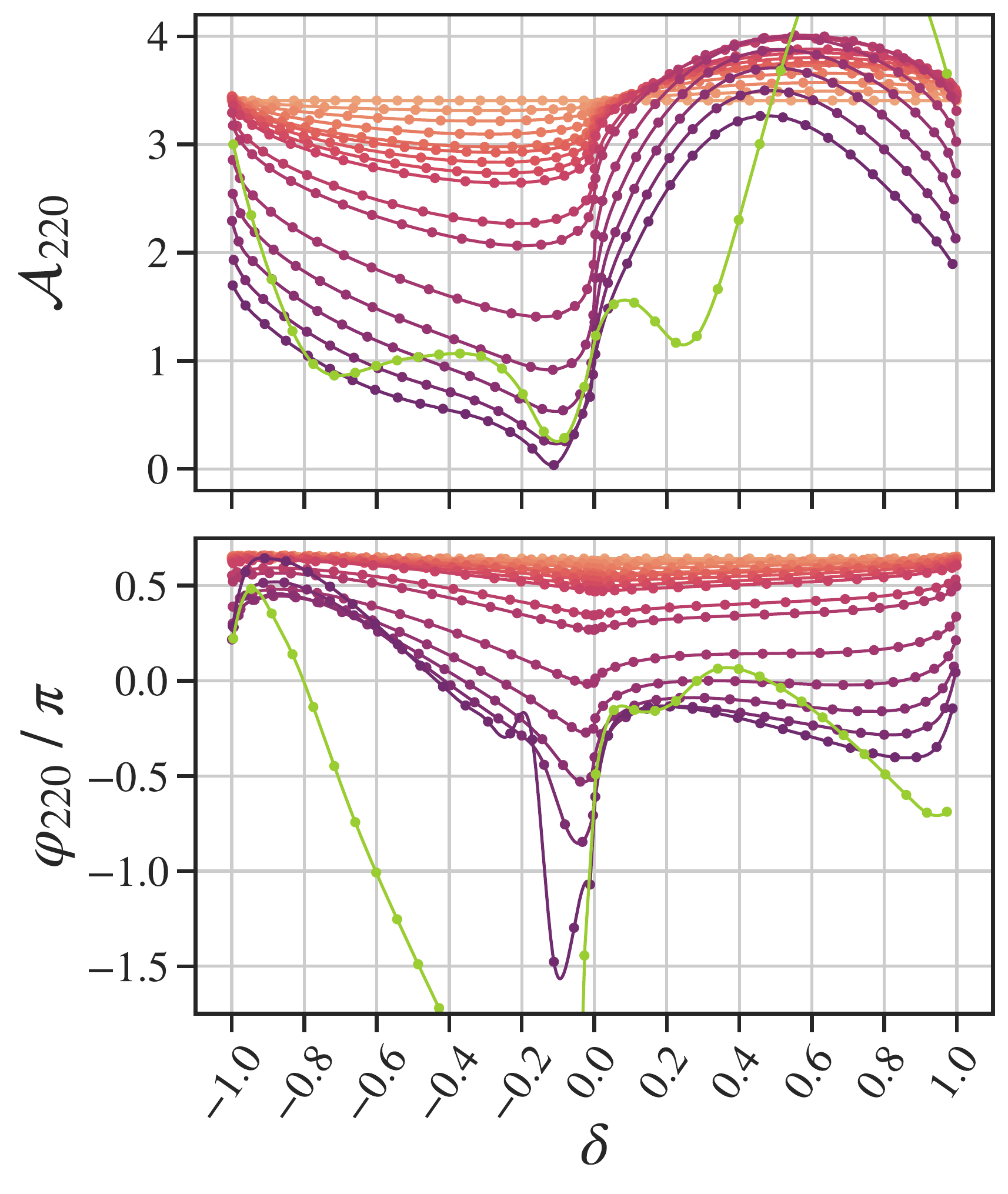}
    \caption{Interpolation of $\mathcal{A}_{220}$ and $\varphi_{220}$ across $\delta$ at fixed $I$ values for $a/M = 0.5$. The prograde calibration data spans $0^\circ \leq I \leq 78^\circ$, which is the left shaded region in Fig.~\ref{fig:grid}. The colors signify the misalignment, with darker colors representing more misaligned inclinations: the $I = 0^\circ$ fits are plotted with the brightest orange and the $I = 78^\circ$ fits are plotted with the darkest purple. Fits from the most misaligned waveform with $I = 86^\circ$ (green) are not included in the calibration data, due to larger numerical errors in the waveforms. }
    \label{fig:interp_22} 
\end{figure}

\begin{figure}
    \centering
    \includegraphics[width=.5\textwidth]{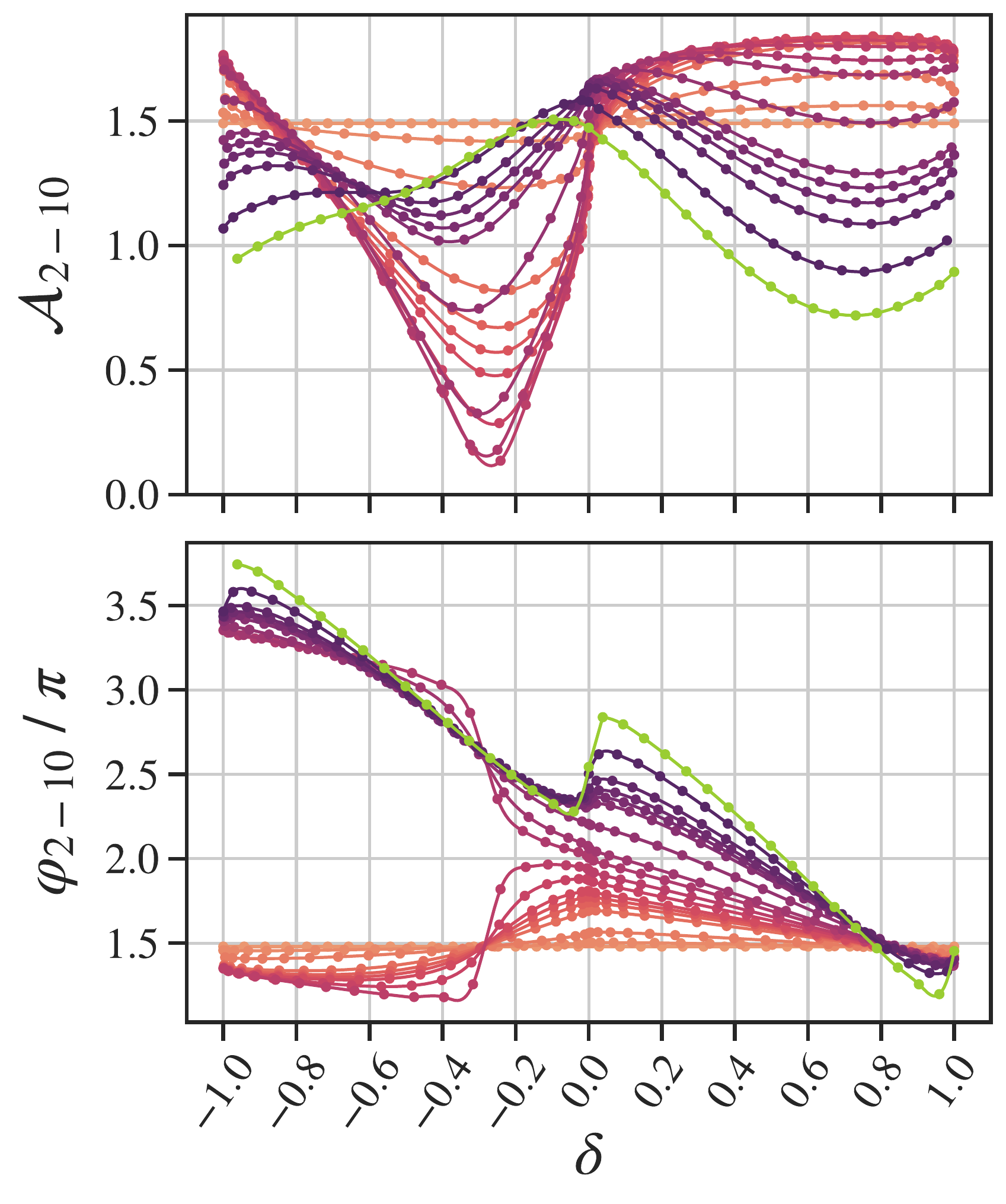}
    \caption{Interpolation of $\mathcal{A}_{2-10}$ and $\varphi_{2-10}$ across $\delta$ at fixed $I$ values for $a/M = 0.5$. The retrograde calibration data spans $101^\circ \leq I \leq 180^\circ$, which is the right shaded region in Fig.~\ref{fig:grid}. The colors signify the misalignment, with darker colors representing more misaligned inclinations: the $I = 180^\circ$ fits are plotted with the brightest orange and the $I = 101^\circ$ fits are plotted with the darkest purple.  Although the most misaligned orbit $I = 93^\circ$ (green) appears well behaved, we do not it include in the calibration data as the fits for other modes at $I = 93^\circ$ appear unreliable.}
    \label{fig:interp_21} 
\end{figure}

Our interpolation procedure consists of sequentially interpolating across each plunge parameter in one dimension. For each interpolation, we use the \texttt{UnivariateSpline} fit routine without smoothing as implemented in \texttt{scipy}. As an example, suppose we want the mode excitation at $(I_0,\delta_0)$, which is not one of the calibration waveforms in Fig.~\ref{fig:grid}. First, we interpolate across $\delta$ to find the mode excitation at $(I,\delta_0)$ at each of the calibration inclinations, while considering potential discontinuities in $d \mathcal{A}_{\ell m n} / d \delta$. The mode amplitudes form closed curves when plotted in $(\theta_{\rm fin},\mathcal{A}_{\ell m n})$ space, joining the $\dot{\theta}_{\rm fin} >0$ and $\dot{\theta}_{\rm fin} < 0$ branches \cite{Lim2019}. Unwrapping this curve and expressing $\mathcal{A}_{\ell m n}$ as a continuous function over $-1 \leq \delta  \leq 1$ can cause $d \mathcal{A}_{\ell m n} / d \delta$ to be extremely large near $\delta = 0$. Therefore, we separately fit two piecewise cubic splines to calibration data for $\delta < 0$ and $\delta > 0$, respectively, and then continuously join these two cubic splines using a linear spline crossing $\delta = 0$. Second, we interpolate across $I$ at fixed $\delta_0$ using a cubic spline to find the mode excitation at $(I_0,\delta_0)$. 

To interpolate the mode phases across simulations, we must align the frame rotation. We perform this alignment by finding the final azimuthal coordinate of the plunge worldline, which approaches a constant in the frame co-rotating with the black hole's horizon. This co-rotating frame rotates relative to the observer frame with angular velocity
\begin{equation}
    d\phi/dt \equiv \Omega_{\rm H} = a/(2Mr_+),
\end{equation}
where $r_+ = M + \sqrt{M^2 - a^2}$ is the coordinate radius of the event horizon. We then rotate each simulation so that the new final azimuthal coordinate of each plunge in the co-rotating frame is uniform across all simulations. 

In Fig.~\ref{fig:interp_22}, we show the interpolation of $\mathcal{A}_{220}$ and $\varphi_{220}$ across $\delta$ for prograde inclinations and $a/M = 0.5$. The line colors correspond to the same inclinations as those plotted in the left shaded region in Fig.~\ref{fig:grid}, where more misaligned orbits are represented with darker colors. We exclude the $I = 86^\circ$ data (green line) when building the final interpolant due to larger numerical errors in the waveforms, which reduce the quality of the fits. At certain points in parameter space, the mode amplitudes may fall below our estimated fit precision (around $\delta \mathcal{A}_{\ell m n} \approx 10^{-2}$). At these points, the phase error tends to be larger. An example of this is the phase discontinuity in the bottom panel of Fig.~\ref{fig:interp_22} that occurs near $\delta = -0.1$ for $I = 78^\circ$ (darkest purple line). We interpolate through these phase discontinuities and find that they do not impact the overall effectiveness of the template because the mode excitation is necessarily small. We expect that increasing the density and quality of calibration waveforms should help remove this type of discontinuity.

In Fig.~\ref{fig:interp_21}, we show the interpolation of $\mathcal{A}_{2-10}$ and $\varphi_{2-10}$ across $\delta$ for retrograde inclinations for $a/M = 0.5$. The mode excitation has periodic boundary conditions in $\delta$, which means $\mathcal{C}_{\ell m n} = \mathcal{A}_{\ell m n} e^{i \varphi_{\ell m n}}$ should be continuous at $\delta = \pm 1$. The mode phases $\varphi_{\ell m n}(I,\delta)$ can satisfy this in two ways, by either ``librating" with $\delta$,
\begin{equation}
    \varphi_{\ell m n }(I,-1) = \varphi_{\ell m n}(I,1),
\end{equation}
or ``circulating" with $\delta$,
\begin{equation}
    \varphi_{\ell m n}(I,-1) = \varphi_{\ell m n}(I,1) + 2\pi k,
\end{equation}
where $k$ is some non-zero integer. When $\ell \neq |m|$, we observe that the mode phases exhibit both libration and circulation, which implies a discontinuity at the transition between these behaviors. For $\varphi_{2-10}$, this phase transition happens between the two calibration inclinations $I = 0.7848 \pi$ and $I = 0.8117\pi$: libration occurs in more aligned plunges with $0.8117 \pi \leq I \leq \pi$, whereas $k = 1$ circulation occurs in more misaligned plunges with $I \leq  0.7848 \pi$. Our ability to further resolve the phase transition point is currently limited by the density of available calibration data. Nevertheless, increasing the density and quality of calibration data will not remove this type of discontinuity.

\begin{figure}
    \centering
    \includegraphics[width=.5\textwidth]{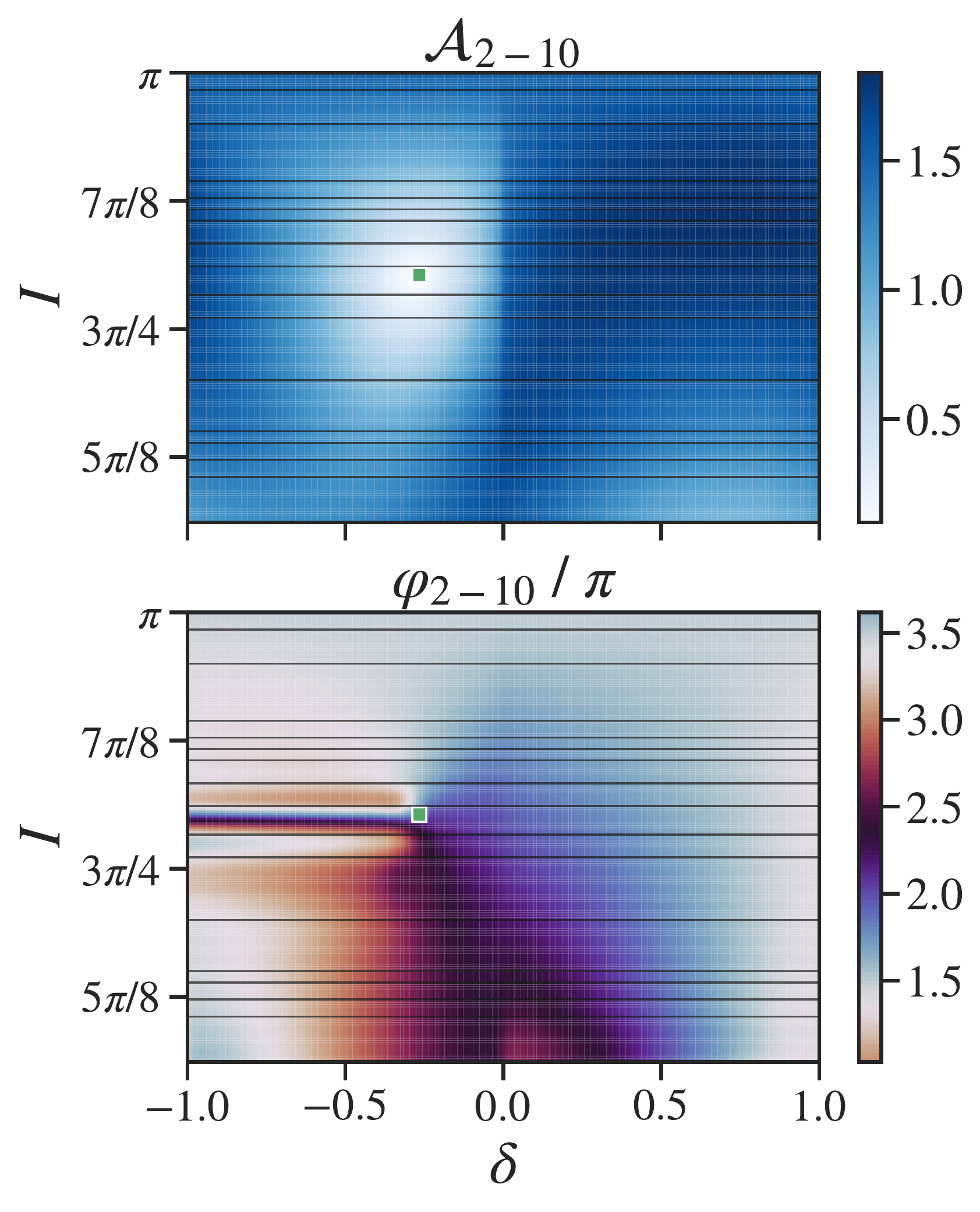}
    \caption{Contour map of the $A_{2-10}$ and $\varphi_{2-10}$ interpolants for retrograde orientations $101^\circ \leq I \leq 180^\circ$ for $a/M = 0.5$. The horizontal lines mark the inclination values of calibration data used to build the interpolants. The color map of the phase plot (bottom) is cyclic with period $2 \pi$. Near $\delta \approx 0$, the ordinary mode amplitudes tend to vary rapidly with $\delta$, which can also be seen in Figs.~\ref{fig:interp_22} and \ref{fig:interp_21}. The green point marks the local minimum of $\mathcal{A}_{2-10}$, at $(I,\delta) = (-0.803\pi,-0.266)$, which is also where the phase transition discontinuity appears.}
    \label{fig:cmaps} 
\end{figure}

In Fig.~\ref{fig:cmaps}, we show a contour of the final interpolant for $\mathcal{A}_{2-10}$ and $\varphi_{2-10}$. The ordinary mode amplitudes tend to vary rapidly with $\delta$ near $\delta = 0$ (which can also be seen in the top panels of Figs.~\ref{fig:interp_22} and \ref{fig:interp_21}), whereas the mirror mode amplitudes tend to vary rapidly with $\delta$ near $\delta = \pm 1$. The local minimum of $\mathcal{A}_{2-10}(I,\delta)$ is located at $(I',\delta') = (0.803\pi,-0.266)$ (marked with a green point). Interestingly, the location of the local minimum is where the phase transition discontinuity occurs. We note that similar behavior is observed in fits in aligned spin NR waveforms, where local minima in the mode amplitudes (as a function of symmetric mass ratio) have also been associated with phase transitions \cite{London2014}. In building our interpolant, there is a choice of where to place the discontinuity. If we enforce continuity in $\varphi_{2-10}$ in the region $\delta > \delta'$ (as illustrated in Fig.~\ref{fig:interp_21}), the discontinuity must be positioned along $I = I'$, $\delta < \delta'$ as shown in Fig.~\ref{fig:cmaps}. Alternately, we can enforce continuity in the region $\delta < \delta'$. This can be achieved by either shifting the phase fits in Fig.~\ref{fig:interp_21} with $I > I'$ by $- 2\pi$, or, by shifting the phase fits with $I < I'$ by $+ 2\pi$. Both would change the position of the discontinuity to be along $I = I'$, $\delta > \delta'$. Our strategy to minimize the effect of the phase transition discontinuity is to align the mode phases such that the discontinuity has the shortest length across the $\delta$ domain. 

\begin{figure}
    \centering
    \includegraphics[width=.5\textwidth]{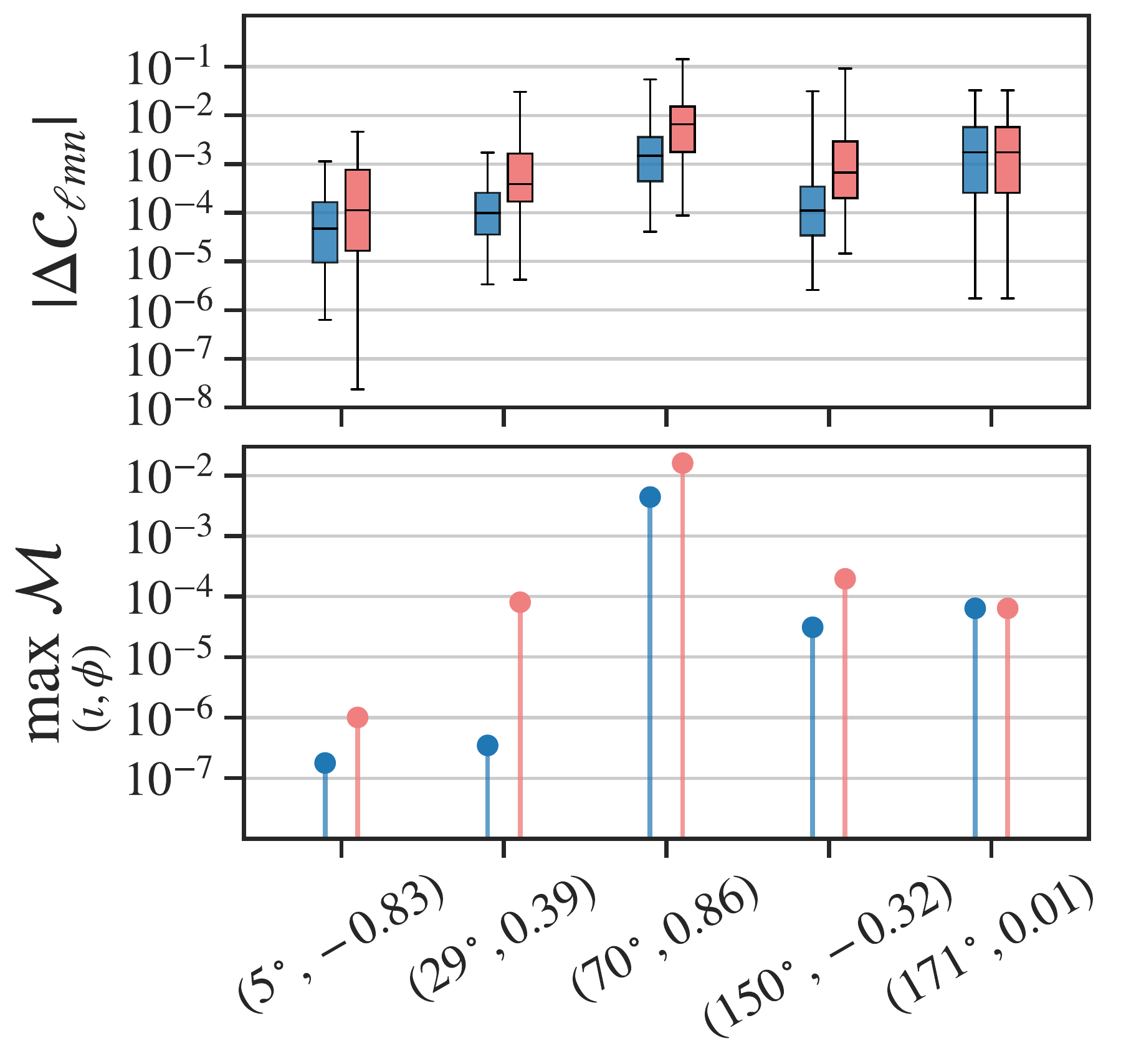}
    \caption{Interpolation error at 5 sample testing points, which we calculate by building a separate testing interpolant that excludes those 5 testing waveforms (testing points circled in Fig.~\ref{fig:grid}). The interpolation error is the difference between the interpolated mode excitation and the mode excitation calculated from direct fits to the testing waveforms. We show the fits that utilize the piecewise cubic splines (blue) along with fits that utilize only linear splines (red). In the top panel we show the 0-25-50-75-100 percentiles in error in $\mathcal{C}_{\ell m n}$ and $\mathcal{C}'_{\ell m n}$ across all modes we fit. We then use the mode amplitudes to construct the ringdown strain [Eq.~(\ref{eq:spheroidaldecomp})] at different sky emission angles $(\iota,\phi)$, and plot the maximum mismsatch across all angles.}
    \label{fig:interp_error} 
\end{figure}

To estimate the error introduced by our interpolants, we build a separate interpolant (``test interpolant") which interpolates through all fits except those from 5 chosen testing waveforms, which are plotted with open circles in Fig.~\ref{fig:grid}. We then measure how well the test interpolant performs at predicting the QNM excitation at the testing points compared to direct fits of the testing waveforms. In the top panel of Fig.~\ref{fig:interp_error} we show the distribution of the magnitude of the difference between the predicted and fitted mode amplitudes,
\begin{eqnarray}
    \Delta \mathcal{C}_{\ell m n} = \mathcal{C}_{\ell m n}^{\rm interpolated} - \mathcal{C}_{\ell m n}^{\rm fitted},
\end{eqnarray}
for all $82 = 42 \times 2$ ordinary and mirror modes that we fit. As a benchmark, we show the interpolation error when using only linear splines. We can improve the error by using cubic splines, which is the method adopted for the rest of this paper. 

In the bottom panel of Fig.~\ref{fig:interp_error}, we plot the maximum mismatch between the strain of the fitted QNM model and interpolated QNM model [Eq.~(\ref{eq:spheroidaldecomp})] from $t = t_0$ to $t = 100M + t_0$ across all emission angles $(\iota, \phi)$, where the mismatch is defined as 
\begin{equation}
    \mathcal{M} = 1- \frac{ |\langle h_1 | h_2 \rangle| }{\sqrt{\langle h_1 | h_1 \rangle \langle h_2 | h_2 \rangle}}.
\end{equation}
As expected, the interpolation error is larger at higher inclinations due to larger numerical error in the waveforms. The interpolation is also worse near $\delta = 0$ where the mode amplitudes vary rapidly with $\delta$. Due to how we sequentially interpolate through the data, these quantities primarily measure the error in incurred during the first interpolation step along $\delta$ at fixed $I$. We expect that the errors incurred during the second interpolation step along $I$ and fixed $\delta$ to be even smaller because the mode amplitudes tend to vary more smoothly with $I$, modulo phase transition discontinuities.

The template is only accurate at late times when the overtone contributions are negligible. Since the onset of ringdown is not defined precisely, we have to be careful not to apply the template to early times where it is invalid, which will lead to biased parameters. Ideally, the overtones should be included to take advantage of additional data closer to the peak. As mentioned in Sec.~\ref{ssec:extraction}, mapping the overtone excitation with sufficient accuracy poses challenges which we leave to future work.

\subsection{\label{sec:bayesian}Parameter estimation}

To estimate how well these templates perform at determining model parameters, we perform Bayesian parameter estimation in the time domain, adding white Gaussian noise with constant variance to the injected waveform.  We justify this approximation by noting that QNM frequencies span a fairly narrow range compared to the full sensitivity of gravitational-wave detectors, though acknowledging that this leaves out important time-domain correlations that would affect ``real-world'' measurements.  We leave a study that incorporates more realistic noise models to future work.

We assume that the noise is fully characterized and adopt a Gaussian likelihood with zero mean and known standard deviation $\sigma$. We define a post-peak SNR,
\begin{eqnarray} \label{eq:SNR}
    \rho_{\rm RD}^2 = \sum_{t_i \geq t_{\rm peak}} \frac{h(t_i)^2}{\sigma^2},
\end{eqnarray}
where the peak of the injected complex strain is
\begin{eqnarray} \label{eq:tpeak}
    t_{\rm peak} = {\rm argmax}\lbrack h_+^2(t) + h_\times^2(t)].
\end{eqnarray}
This definition is convenient in practice because the peak of the signal can be recovered fairly accurately.

The physical parameters required to uniquely label an observed waveform from a quasi-circular, large mass ratio black hole binary are:
\begin{equation} \label{eq:allpars}
    \boldsymbol{\lambda} = \left \lbrace a, M, I, \delta,\iota,\phi,F_+,F_\times,\psi,\mu/d_L \right\rbrace.
\end{equation}
As mentioned in Sec.~\ref{sec:intro}, we assume that the remnant parameters $a$ and $M$ are known prior to estimating the mode amplitudes. This assumption reflects a future detection scenario in which the remnant parameters (and hence, the mode frequencies) are tightly constrained through inspiral-only or IMR models. Throughout the remainder of this paper, we also assume a single detector and fix the antenna response $F_+ = 1, F_\times = 0$ and the polarization angle $\psi =0$.

We now briefly discuss how we choose the masses. In terms of the simulated detector sensitivity, the analysis is independent of the mass since we are using a white noise model.  We choose masses such that the total system mass is compatible with binaries currently being observed, while remaining at least marginally compatible with the requirements of BHPT.  We choose the larger mass to be $M = 100M_\odot$, which is consistent with that of larger black holes observed by the Advanced LIGO and Advanced Virgo detectors. We choose the smaller mass to be $\mu = 1M_\odot$, which is consistent with that of neutron stars.

At a mass ratio of $q \equiv M / \mu = 100$, we must consider the limits of perturbation theory in studying the ringdown. In first-order BHPT, the remnant parameters are the same as the larger BH's and the component masses are just scaling parameters --- $M$ fixes the physical timescale of the system and the GW strain is proportional to $\mu$. At $q = 100$, we can estimate the error introduced by this approximation as the expected change in mass of the larger BH and the remnant BH, which is on the order of $ \Delta M \lesssim 1\%$. Note that this difference is small, and that modeling this variation is more relevant during the merger and early ringdown when the remnant spacetime may still be stabilizing. As our model
only includes the fundamental modes, we do not use early ringdown data and ignore these effects. A more careful consideration of these approximations may be warranted when using our framework to study overtones closer to the merger.

A reference time such as $t_{\rm peak}$ is needed to fix the detector frame. In building our ringdown waveform model, we calculate the fiducial start time $\hat{t}_0(\boldsymbol{\lambda})$ [Eq.~(\ref{eq:lightring})] and the peak strain $\hat{t}_{\rm peak}(\boldsymbol{\lambda})$ [Eq.~(\ref{eq:tpeak})] in the simulation frame. Then, to analyze the strain data, we use the peak time to find the fiducial time in the detector frame,
\begin{equation}
    t_0 = t_{\rm peak} + \lbrack \hat{t}_0 - \hat{t}_{\rm peak} \rbrack,
\end{equation}
where $t_{\rm peak}$ is measured from the data, and which we assume is known precisely.

We prepare the data by down-sampling to 4096Hz and removing points outside a fitting window $\lbrack t_{\rm cut}, t_{\rm end} \rbrack$. The start of the fitting window $t_{\rm cut}$ is chosen to be late enough that we expect the ringdown model to be valid; the details of this choice are given in Sec.~\ref{sec:casestudy}. The data are truncated at $t_{\rm end}$ after which noise dominates the signal. Given the fitting window, we also introduce the post-cut SNR $\rho_{\rm cut}$,
\begin{eqnarray} \label{eq:SNRcut}
    \rho_{\rm cut}^2 = \sum_{t_i \geq t_{\rm cut}} \frac{h(t_i)^2}{\sigma^2},
\end{eqnarray}
which characterizes how loud the post-cut signal is. Since $t_{\rm cut}$ is generally later than $t_{\rm peak}$ for our analyses, the SNR of data actually used for inference is lower than the post-peak ringdown SNR, $\rho_{\rm cut} < \rho_{\rm RD}$.

We summarize the 5 remaining search parameters and their priors below:

\begin{enumerate}
    
    \item[(i)] $I$ --- Spin-orbit misalignment, or inclination, of the binary. We choose a prior which assumes the spin direction is isotropic, or uniform in $\sin(I)$ over $ \lbrack 0,\sin^{-1}(0.98) \rbrack \cup \lbrack \pi - \sin^{-1}(0.98),\pi \rbrack$.  The removed region corresponds to the high misalignments over which our fits are unreliable.
    \item[(ii)] $\delta$ --- Normalized plunge phase parameter, which can be mapped to $\lbrace \theta_{\rm fin},{\rm sgn}(\dot{\theta}_{\rm fin}) \rbrace $ through Eq.~(\ref{eq:delta}). We choose a prior which assumes that the final phase of an orbit is uniform in $-\pi \leq \chi_f \leq \pi$, where the final phase is defined as
    \begin{eqnarray}
        \cos(\chi_f) & = & \cos(\theta_{\rm fin}) / \sin(I), \\
        {\rm sgn}(\chi_f) & = & {\rm sgn}(\dot{\theta}_{\rm fin}).
    \end{eqnarray}
    \item[(iii)] $\iota,\phi$ --- Spherical coordinates describing the angular direction of emission, with $\iota = 0$ aligned with the black hole spin. The priors are uniform in $\cos(\iota)$ and $\phi$. We only sample over $0 \leq \iota \leq \pi/2$ because of reflection symmetry --- sources with $\iota$ and $\pi - \iota$ are degenerate when only using a single polarization $h_+$.
    \item[(iv)] $d_L$ --- Luminosity distance in units of Gpc. {We set the priors uniform in volume, so that $d_L^3$ is uniform between $\lbrack 0, 1 \rbrack$.}  
\end{enumerate}

To sample the posterior probabilities of these parameters, we use the open-source \texttt{Bilby} \cite{Ashton2019Bilby:Astronomy,Romero-Shaw2020BayesianCatalogue} platform which is a wrapper for several Bayesian posterior samplers. We use the \texttt{DYNESTY} sampler as implemented in \texttt{Bilby} which we find is flexible enough to capture some of the degeneracies and other complex features of the posterior that emerge \cite{Speagle2020Dynesty:Evidences}. We use the \texttt{rwalk} sample method and default settings, but set the minimum number of walks to 25, and set the number of live points at 5000 for the analyses in Secs.~\ref{sec:casestudy} and \ref{sec:retrograde}, at $2000N$ for each $N$ mode model in Sec.~\ref{sec:Nmode}.

We also use our Bayesian framework to perform a source-agnostic analysis, where the relative mode amplitudes are not constrained by informative priors. There are $2N$ search parameters for $N$ included modes, assuming the mode frequencies are known and only a single polarization. As more modes are included, the sampling becomes increasingly degenerate and computationally expensive. In order to determine the number and variety of excited modes supported by the data we perform Bayesian model selection. Recently, Ref.~\cite{Bustillo2020} performed such a model selection and calculated the statistical evidence for including the $(2,2,\leq 3)$ overtones in describing GW150914-like signals. We follow a similar approach and compare models with different sets of angular harmonics. However, the angular harmonics are not ordered by damping times and the variety of excited modes varies significantly across plunge geometries. As a result, the model space we consider is larger, but can be made more tractable through an algorithm (greedy or otherwise), prior information, or both.

The Bayesian evidence of a model is defined as
\begin{equation}
    \mathcal{Z} = \int d\boldsymbol{\lambda} \mathcal{L}(d | \boldsymbol{\lambda}) \pi(\boldsymbol{\lambda}),
\end{equation}
where $\mathcal{L}(d|\boldsymbol{\lambda})$ is the likelihood function describing observed data $d$ given model parameters $\boldsymbol{\lambda}$, and $\pi(\boldsymbol{\lambda})$ is the prior distribution. Given two models $A$ and $B$, the relative probability of model $A$ over model $B$ is ${Z}_A / {Z}_B$. The threshold is usually expressed in terms of the log Bayes factor, where $\log \mathcal{B} =  \log({Z}_A / Z_B) \simeq 5$ is a common choice signifying that model A is roughly 150 times more likely than model B. Importantly, these models may include different numbers of parameters, and the Bayes factor inherently penalizes additional degrees of freedom that do not contribute to the quality of the fit. We use the Bayesian evidence to find the optimal ringdown model across both the number and variety of modes included in the model.

\section{\label{sec:results}Results}

We now discuss the results we obtain measuring mode amplitudes with our Bayesian inference framework. We restate the primary waveform model which we are using to perform the parameter estimation, which is a function of 5 search parameters [the remaining of which are fixed in Eq.~(\ref{eq:allpars})]:
\begin{align}\label{eq:RDmodel}
&h_+^{\rm RD}(t;I,\delta,\iota,\phi,d_L) =
\nonumber  \\
& \frac{\mu}{d_L} \mathfrak{R} \bigg\lbrace \sum_{(\ell,m) \in \mathcal{S}} \Big[  \mathcal{A}_{\ell m0} e^{-i[\sigma_{\ell m0}(t-t_0) - \varphi_{\ell m0}]}  {_{-2}}S^{a\sigma_{\ell m0}}_{\ell m0}(\iota,\phi)
\nonumber\\ 
& + \mathcal{A}'_{\ell m0} e^{-i[\sigma'_{\ell m0}(t-t_0)-\varphi'_{\ell m0}]}  {_{-2}}S^{a\sigma'_{\ell m0}}_{\ell m0}(\iota,\phi)\Big] \bigg \rbrace\;.
\end{align}
For the remainder of the paper, we drop the polarization subscript ``+" as we only consider one polarization. The BH parameters $a$ and $M$ and the QNM frequencies $\sigma_{\ell m0} = \sigma_{\ell m0}(a,M)$ are considered known and fixed to their injection values. We also fix $\mu$ since it is degenerate with the distance $d_L$. With the spin fixed, the spheroidal harmonics ${_{-2}}S^{a\sigma_{\ell m0}}_{\ell m0}$ are just functions of the the emission angles $\iota$ and $\phi$. The set of modeled QNMs $\mathcal{S}$ includes all fundamental modes up to $\ell \leq 8$ and $\ell - |m| \leq 4$. Finally, the QNM amplitudes and phases $\mathcal{A}_{\ell m0}$, $\mathcal{A}'_{\ell m0}$, $\varphi_{\ell m0}$, and $\varphi'_{\ell m0}$ along with the fiducial time $t_0$ are treated as functions of the plunge parameters $I$ and $\delta$. This represents our prior knowledge of how modes are excited as a function of plunge geometry.

\subsection{\label{sec:casestudy} Measuring mode amplitudes with a source model for excitation}

\begin{table}[b]
\caption{\label{tab:table1}
Intrinsic and extrinsic mode amplitudes of the 8 most dominant modes for an misaligned system with $I = 137^\circ$ and line-of-sight inclination $\iota = \pi/3$. Additional system parameter defined at the beginning of Sec.~\ref{sec:casestudy}. The extrinsic mode amplitudes are defined in Eq.~(\ref{eq:extrinsic}).}.
\begin{ruledtabular}
\begin{tabular}{cccc}
& $\mathcal{A}_{\ell m 0}$ & $\mathcal{A}'_{\ell -m 0}$  & $C_{\ell m 0} (d_L / \mu)$ \\
\colrule
(2,1)  & 1.69 & 0.09 & 0.72 \\
(2,-2) & 1.18 & 1.71 & 0.67 \\
(2,0)  & 1.93 & 0.49 & 0.54 \\
(2,-1) & 1.74 & 1.04 & 0.39 \\
(2,2)  & 1.00 & 0.04 & 0.35 \\
(3,-3) & 0.39 & 0.61 & 0.29 \\
(3,-1) & 0.69 & 0.24 & 0.22 \\
(3,0)  & 0.62 & 0.09 & 0.21
\end{tabular}
\end{ruledtabular}
\end{table}

We focus our attention on analyzing the ringdown from a highly misaligned plunge, which excites modes beyond the ``standard set". We inject one of our Teukolsky waveforms with source parameters $a/M = 0.5$, $\mu = 1M_\odot$, $M = 100M_\odot$, $I = 137^\circ = 2.39$, $\delta = 0.38$, $d_L = 0.5\ {\rm Gpc}$, $\iota = \pi/3$, and $\phi = 5 \pi / 4$. At a spin-orbit misalignment of $I = 137^\circ$ and line-of-sight inclination of $\iota = \pi/3$, the ringdown spectrum is relatively complicated. In Table I, we show the amplitudes of the 8 most dominant modes in the ringdown, ranked by their extrinsic amplitudes $C_{\ell m}$ which we define as
\begin{equation}
    C_{\ell m 0} = \sqrt{B_{\ell m 0}^2 + {B'}_{\ell -m 0}^2},
\end{equation}
where
\begin{eqnarray} \label{eq:extrinsic}
    B_{\ell m 0}  & = & \frac{\mu}{d_L} \mathcal{A}_{\ell m 0}   \left|{_{-2}}S^{a\sigma_{\ell m 0}}_{\ell m 0}(\iota,\phi)\right| \nonumber \\
    B'_{\ell m 0} & = & \frac{\mu}{d_L} \mathcal{A}'_{\ell m 0}  \left|{_{-2}}S^{a\sigma'_{\ell m 0}}_{\ell m 0}(\iota,\phi)\right|.
\end{eqnarray}
The quantities $C_{\ell |m|}$ and  $C_{\ell -|m|}$ are the total prograde and retrograde amplitudes of the $(\ell, \pm m)$ modes, respectively, calculated as the quadrature sum of the relevant ordinary and mirror modes.

For the injection waveform we consider here, we find the retrograde $(2,-2)$ and prograde $(2,1)$ modes are strongest (Table~\ref{tab:table1}). This broad excitation of both prograde and retrograde modes is characteristic of highly misaligned systems, occurring across a wide range of plunge angles $90^\circ  \lesssim I \lesssim  160^\circ$ and all spin cases we tested in the range $0.1 \leq a/M \leq 0.9$.  Although the mode excitation for comparable mass binaries has not been completely mapped, recent precessing waveform models have demonstrated similar behavior in the ringdown. For example, in Fig.~11 in Ref.~\cite{Hamilton2021}, both the prograde and retrograde mode frequencies are present in the $\ell = |m| = 2$ spherical mode when the effective spin is highly misalinged with the orbital angular momentum over a range of misalignments quite similar to what we observe in the large mass ratio limit, $90^\circ \lesssim \theta_{\rm LS} \lesssim 150^\circ$. Further work is needed to  characterize the mode excitation in these waveforms and understand its connection with our findings in the large mass ratio regime.

\begin{figure}
    \centering
    \includegraphics[width=.5\textwidth]{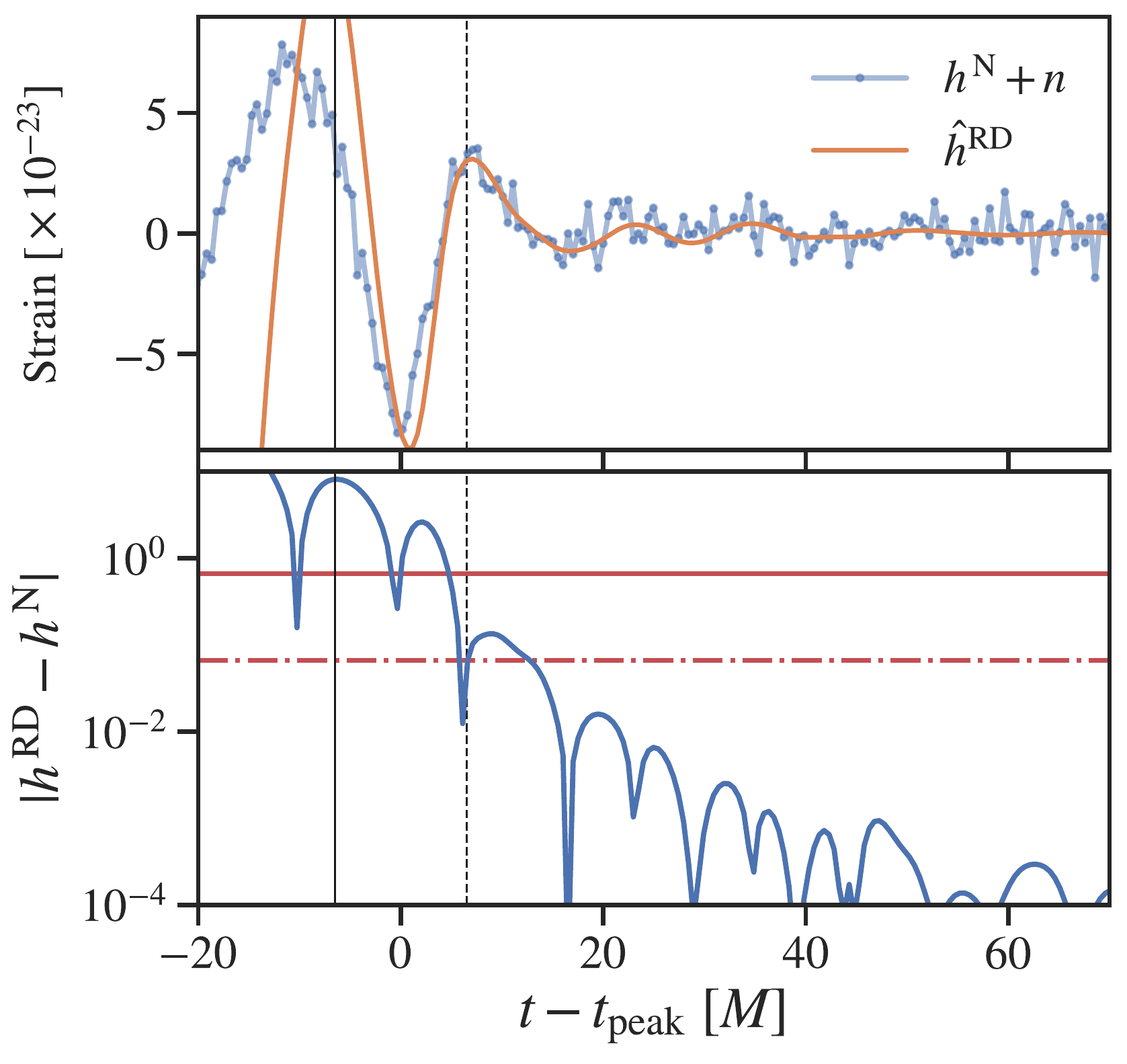}
    \caption{The injected waveform with additive Gaussian noise $h^{\rm N}(t) + n(t,\sigma)$ used in our parameter estimation, along with a model of the fundamental QNM content, $\hat{h}^{\rm RD}$. The time is plotted relative to $t_{\rm peak}$. The light ring crossing time occurs at $t_0 = -6.5M$ (solid black line), and the analysis start time is $t_{\rm cut} = 6.5M$ (dashed black line). The noise level is set at $\sigma = 0.64$ (solid red line) which corresponds to a post-peak SNR of $\rho_{\rm RD} = 25$ for data sampled at 4096 Hz. The $\sigma / 10$ level is also shown (dot-dashed red line). }
    \label{fig:strain_residual} 
\end{figure}

In the top panel of Fig.~\ref{fig:strain_residual}, we plot the injection waveform with added Gaussian noise $h^{\rm N}(t) + n(t,\sigma)$ when the post-peak SNR is $\rho_{\rm RD} = 25$. {For comparison, with Advanced LIGO/Virgo, an event as loud as GW150914 (with a final mass $M \approx 70\ M_\odot$) has a post-peak SNR of $\rho_{\rm RD} = 15$ \cite{Bustillo2020}. Future detectors such as Einstein Telescope and Cosmic Explorer will routinely achieve $\rho_{\rm RD} \gtrsim 30$ \cite{Bhagwat2020,Berti2016}.} We also plot the QNM content $\hat{h}^{\rm RD}$ which is the waveform model [Eq.~(\ref{eq:RDmodel})] evaluated with the source parameters corresponding to the injection waveform. In the bottom panel of Fig.~\ref{fig:strain_residual}, we plot the magnitude of the residual between the injection waveform and the QNM model, $|\hat{h}^{\rm RD} - h^{\rm N}|$. At early times, the data is not well described by a superposition of fundamental QNMs, so including this portion of the data will bias the parameter estimation. We therefore analyze the segment of data between $t_{\rm cut} = 6.5M + t_{\rm peak}$, after which the residual power is consistent with noise and falls below the $\sigma/10$ level, and $t_{\rm end} = 75M + t_{\rm peak}$, after which the signal is dominated by noise. The SNR of the signal after truncation is $\rho_{\rm cut} = 11.2$.

\begin{figure}
    \includegraphics[width=.45\textwidth]{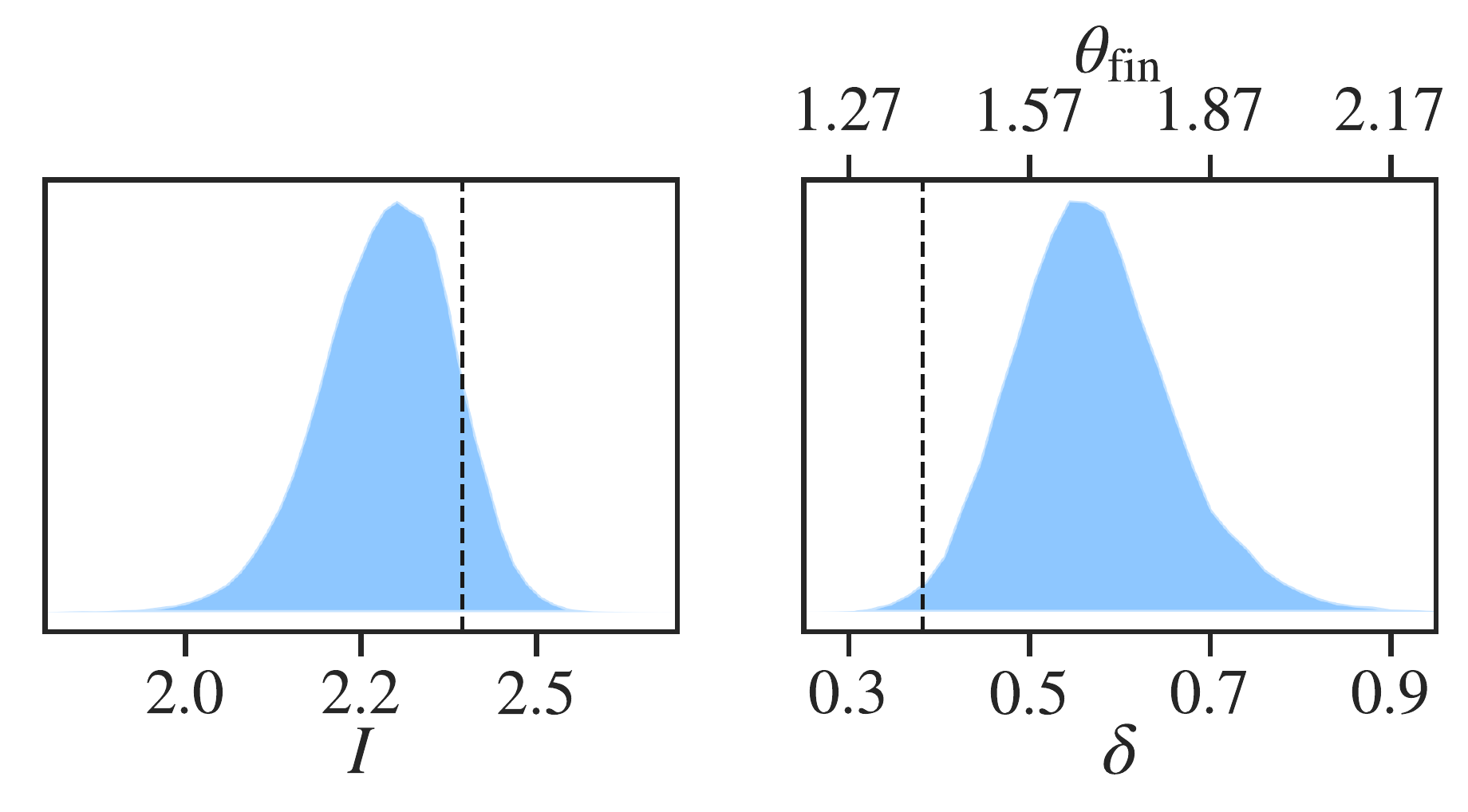}
    \includegraphics[width=.45\textwidth]{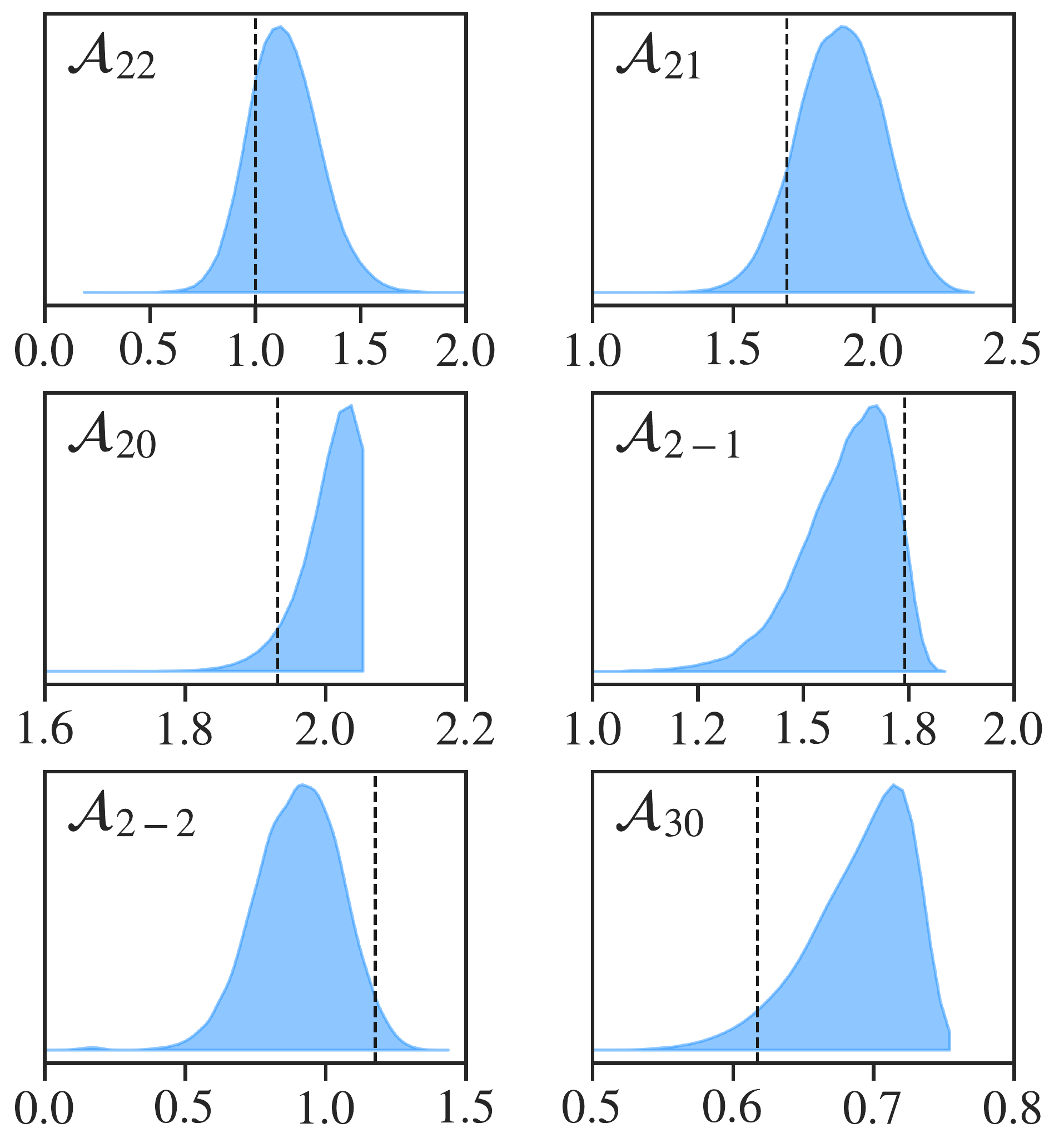}
    \caption{Posterior distributions of the plunge parameters $I$ and $\delta$ and several mode amplitudes. We smooth the posterior distributions using a Gaussian KDE. The dashed line indicate the injected values. The posteriors for the $(2,0,0)$, $(2,-1,0)$, and $(3,0,0)$ modes show a steep cutoff, which is due to the prior probability going to zero. For example, the global maximum for the $(2,0,0)$ mode amplitude across plunge parameters is $\max_{(I,\delta)} \mathcal{A}_{200} = 2.05$, above which the prior (and posterior) vanishes. }
    \label{fig:thinc_delta} 
\end{figure}

\begin{figure}
    \centering
    \includegraphics[width=.45\textwidth]{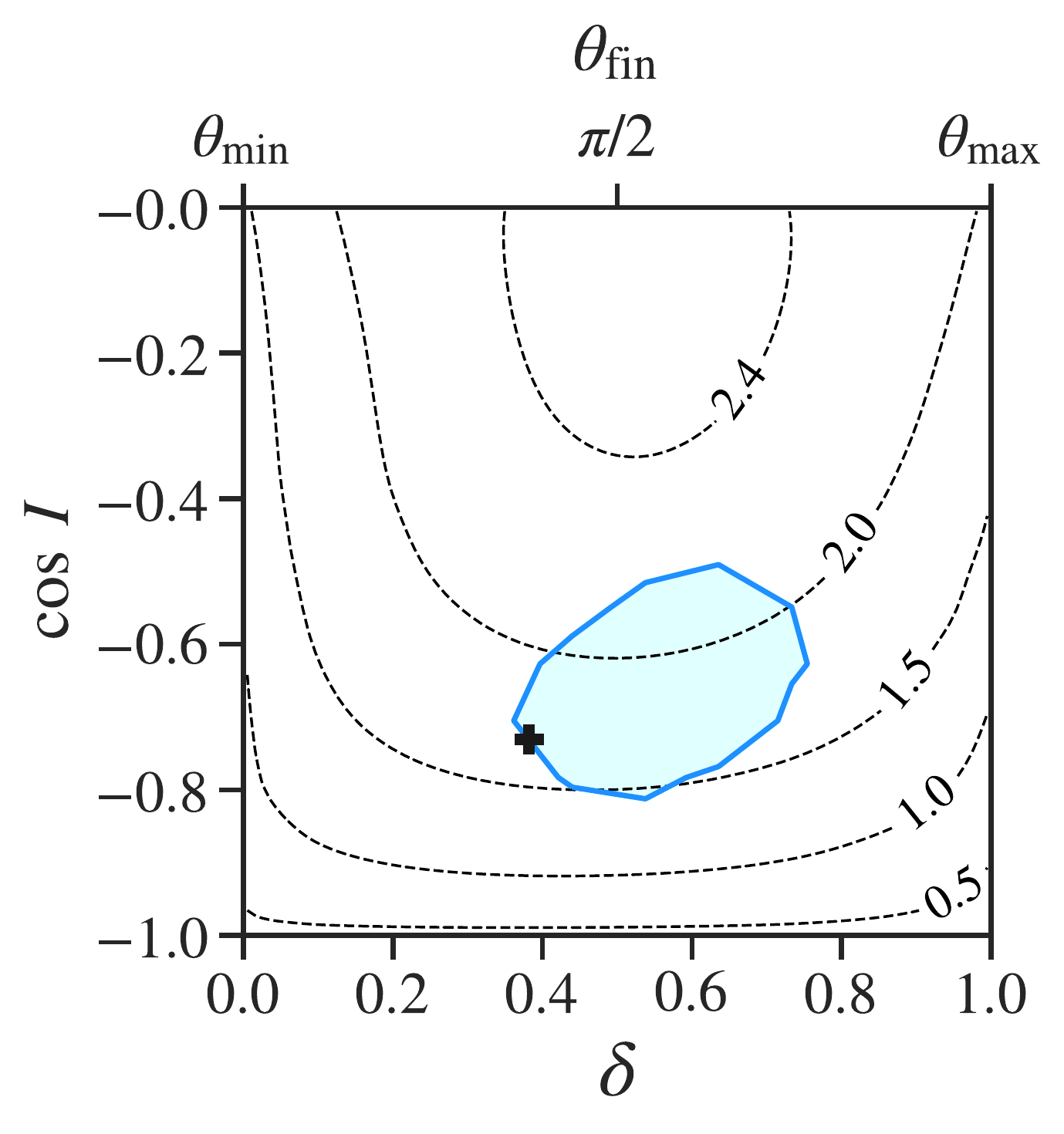}
    \caption{$2$-$\sigma$ contour of the marginalized posterior distribution of $\mathcal{A}_{210}$ (shaded blue) as a function of the plunge parameters. The black marker is the injected value, and the contour lines are levels of constant $\mathcal{A}_{210}$. Plunges that terminate near the turning points of the orbit have $\theta_{\rm fin} = \theta_{\rm min}$ or $\theta_{\rm fin} = \theta_{\rm max}$. Equatorial plunges have $\cos(I) = \pm 1$ and $\theta_{\rm min} = \theta_{\rm max}$.}
    \label{fig:landscape} 
\end{figure}

In the top panel of Fig.~\ref{fig:thinc_delta}, we show the marginalized posterior distributions for the plunge angles $I$ and $\delta$. We post-process the posterior samples in the plunge angles and map them to posterior samples in the mode amplitudes, shown in the bottom panels of Fig.~\ref{fig:thinc_delta}. Several features exhibited by this particular case apply to all plunge geometries and spins. For instance, the inclination $I$ is more tightly constrained than the plunge phase $\delta$, and the bias in the estimated plunge phase $\delta$ is large compared to the bias in the estimated mode amplitudes. For $I$ and $\delta$, respectively, the size of the 90\% credible interval relative to the median is 13.3\% and 55.3\%, and relative to the prior range is 9.9\% and 14.6\%. The relative difficulty in constraining $\delta$ is due to the behavior of the mode amplitudes across the plunge angles, which we illustrate in Fig.~\ref{fig:landscape} with the dominant mode $(\ell,m,n) = (2,1,0)$. The contours of constant amplitude show the mode amplitude is more sensitive to $I$ near $\delta \approx 0.5$. The exception is when the plunge terminates near a turning point in its orbit when $\delta \approx 0,\pm 1$, or $\theta_{\rm fin} = \theta_{\rm min},\theta_{\rm max}$, in which case the mode excitation is highly sensitive to $\delta$. At these turning points, $\delta$ can be tightly constrained, modulo a degeneracy at $\delta = 0$ and $\delta = \pm 1$ due to the reflection symmetry of Kerr [c.f. Eq.~(4.6) in LKAH]. As the inclination goes to $I = 0$ or $I = \pi$, the polar range of motion decreases and the dependency on $\delta$ disappears.

\subsection{\label{sec:Nmode} Source-agnostic analysis}

We now analyze the same injection waveform using a secondary source-agnostic model that leaves the mode amplitudes and phases unconstrained. This general $N$-mode model can be written
\begin{eqnarray} \label{eq:Nmode}
    h^{\rm RD}_{N,M}  & = & \sum_{k \in M} \left\lbrack F_k \cos(\omega_k t)  + G_k \sin(\omega_k t) \right\rbrack e^{-\frac{t-t_0}{\tau_k}},\qquad  \\
    M  & = & \lbrace (\ell_1,m_1), (\ell_2,m_2), ..., (\ell_N, m_N) \rbrace,
\end{eqnarray} 
where $k$ labels each mode and $\omega_k$ and $\tau_k$ are the (positive) real and imaginary components of the mode frequency. Assuming that the remnant mass and spin are known and a single polarization, there are $2N$ fitting parameters. The set of modes to model, $M$, is a hyperparameter. The fitting parameters $F_k$ and $G_k$ are completely degenerate with the fiducial time $t_0$ due to time shift symmetry, so we set $t_0$ to be the same as the injection value for our source-informed analysis for ease of comparison. The spheroidal harmonics and mode amplitudes are nuisance parameters, and can be related to the fitting parameters $F_{\ell m}$ and $G_{\ell m}$ by
\begin{eqnarray}\label{eq:fitpars}
    F_{\ell m} & = & \frac{\mu}{d_L} \mathfrak{R}\Big \lbrace \mathcal{A}_{\ell m 0}e^{i \varphi_{\ell m 0}} {_{-2}}S^{a\sigma_{\ell m 0}}_{\ell m 0}(\iota,\phi) \qquad \nonumber \\
     & & \quad + \mathcal{A}'_{\ell -m 0} e^{i \varphi'_{\ell -m 0}} {_{-2}}S^{a\sigma'_{\ell -m 0}}_{\ell -m 0}(\iota,\phi) \Big \rbrace, \\
    G_{\ell m} & = & \frac{\mu}{d_L} \mathfrak{I}\Big \lbrace - \mathcal{A}_{\ell m 0}e^{i \varphi_{\ell m 0}} {_{-2}}S^{a\sigma_{\ell m 0}}_{\ell m 0}(\iota,\phi) \qquad \nonumber \\
     & & \quad + \mathcal{A}'_{\ell -m 0} e^{i \varphi'_{\ell -m 0}} {_{-2}}S^{a\sigma'_{\ell -m 0}}_{\ell -m 0}(\iota,\phi) \Big \rbrace.
\end{eqnarray}

Performing Bayesian model selection with non-precessing ringdowns is fairly straightforward because the model space is small. The assumed set of modes is limited to either the $\ell = |m| = 2$ prograde or retrograde modes \cite{Isi2019,Bustillo2020}. Testing for the presence of an additional mode entails adding the next higher overtone. However, when considering multiple excited angular modes, the model space is much larger. The number of unique models to select from is naively
\begin{equation}
    N_M = \sum\limits_{N=1}^{N_{\rm max}} \binom{N_{\rm modes}}{N},
\end{equation}
where $N_{\rm modes}$ is the number of modes that might be in the ringdown and $N_{\rm max}$ is a maximum number of modes to include in any given model. In a detection scenario, $N_{\rm max}$ is practically bounded since only the loudest modes should be measurable. The Bayesian evidence should stay the same, or decrease, if additional modes are included but cannot be constrained by the data. If we restrict the possible modes to $2 \leq \ell \leq 4$, $n=0$ and limit the search to $N_{\rm max} = 6$, then there are still $N_M \approx 10^5$ models to compare.

Partial knowledge about the source (such as constraints on the mass ratio or spin alignment) should be employed to target a subset of likely modes. For instance, at comparable mass ratios and aligned spins, the spectrum converges to the standard set of modes with $\ell \lesssim 5$. For this analysis, we restrict the hyperparameter space to the loudest $N_{\rm max} = 6$ modes in Table.~\ref{tab:table1}, which we denote as a set containing six modes,
\begin{equation}
    M_0 = \lbrace (2,1), (2,-2), (2,0), (2,-1), (2,2), (3,-3) \rbrace.
\end{equation}
This makes performing Bayesian model selection more tractable for our study, but can also be considered as some partial knowledge of the source. 

In addition to limiting the set of modes to search over, we implement a greedy algorithm, representing each model as a unique subset of $M_0$. We start with $N = 0$ modes, and add one additional mode at each iteration to find the model with maximum Bayesian evidence. The algorithm is as follows:
\begin{enumerate}
\item[(i)] Specify $M_0$ which is the set of possible modes to search over. Initialize set of ``final" models $S = \lbrace \lbrace \rbrace \rbrace$, only containing the empty set ($N = 0$ model). Initialize $N = 1$.
\item[(ii)] Calculate $S_{N-1}$ which is the subset containing all models in $S$ with $N-1$ modes. Calculate $T_N$ which is the set containing all possible ways to add one of the modes in $M_0$ to one of the models in $S_{N-1}$.
\item[(iii)] Add $T_N$ to $S$ and calculate the maximum Bayesian evidence $\mathcal{Z}_{\rm max}$ across all models in $S$. Remove all models in $S$ with insufficient evidence, $\log(\mathcal{Z}/\mathcal{Z}_{\rm max}) < - \beta$.
\item[(iv)] Stop if $N = N_{\rm max}$. Otherwise iterate from (ii) with $N = N + 1$. 
\end{enumerate}
At each iteration, we consider all possible models with one additional mode, drawing from the assumed set of potential modes $M_0$. Then, we calculate the evidence for all the models, and drop the models which have relatively little support compared to highest evidence model, continuing onto the next iteration. We set the log evidence threshold for a model to be dropped at $\beta = 5$, and converge on a final set of models $S$. {We set the priors for all the mode amplitudes to be uniform in the interval $\lbrack -1000,1000 \rbrack$, which is wide enough so that the posteriors are not truncated by the bounds of our priors (with exception to the $N = 6$ model, which exhibits degeneracies which cause the posteriors to extend beyond $|F_{\ell m}|,|G_{\ell m}| > 2000$).}

\begin{table}[t!]
\caption{\label{tab:greedy}
Result of the greedy algorithm described in the text, which returns a final set of ringdown models with the largest Bayesian evidences. We set the maximum number of modes in any considered model to be $N_{\rm max} = 6$ and only search for modes in the set $M_0 = \lbrace (2,1), (2,-2), (2,0), (2,-1), (2,2), (3,-3) \rbrace$. The maximum evidence was found for a two-mode model, with a signal-to-noise log Bayes factor $\log B_{\rm max} = 59.0$. Setting our threshold parameter at $\beta = 5$, we drop all models with a Bayes factor lower than $54.0$, which excludes some models with $N = 1,2$ and all models with $N \geq 3$. The maximum estimated error in $\log B$ across all runs is $0.082$.}
\begin{ruledtabular}
\begin{tabular}{ l | l}
 $M$ &  $\log B$ \\ \hline 
(2,-2) & 55.8 \\
 \hline
(2,-1),(2,-2) & \textbf{59.0} \\ 
(2,0),(2,-1) & 57.4 \\ 
(2,0),(2,-2) & 57.2 \\ 
(2,1),(2,-2) & 55.2 \\ 
(2,1),(2,-1) & 54.2 \\ 
\end{tabular}
\end{ruledtabular}
\end{table}

\begin{figure}
    \centering
    \includegraphics[width=.5\textwidth]{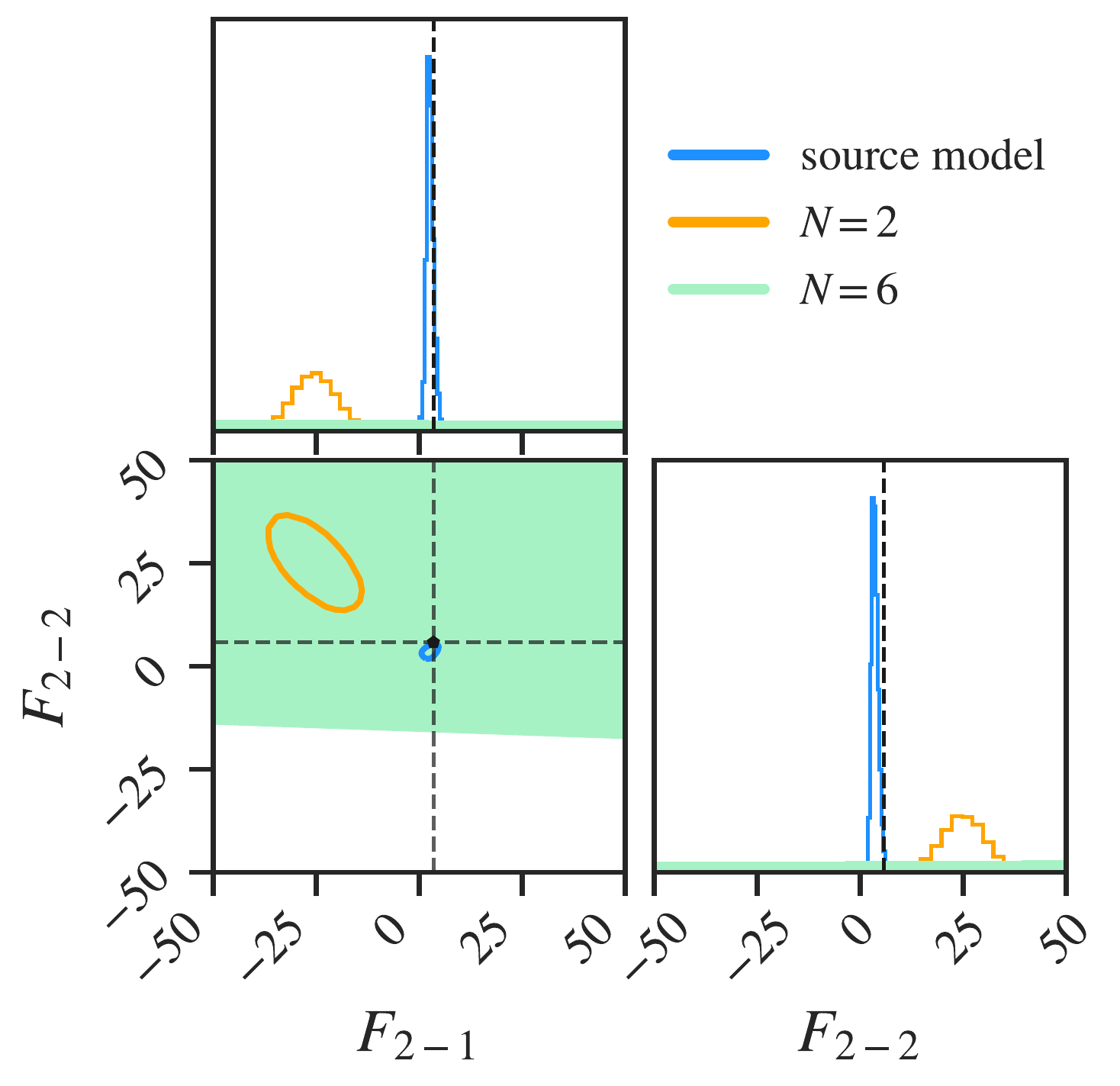}
    \caption{The recovered fitting amplitudes in a source-agnostic measurement. We use the same injection waveform, noise, and fitting window as in Sec.~\ref{sec:casestudy}, and plot the posterior distributions for the $F_{2-1}$ and $F_{2-2}$ fitting amplitudes and 2-$\sigma$ contours. We show the model with the greatest Bayesian evidence, $M = \lbrace (2,-1),(2,-2) \rbrace$ (yellow), along with the $N = 6$ model, $M = \lbrace (2,2),(2,1),(2,0),(2,-1),(2,-2),(3,-3) \rbrace$ (green). We compare these results with the posteriors from the source-informed analysis in Sec.~\ref{sec:casestudy}, which constrained the mode excitation as a function of the plunge parameters (blue). }
    \label{fig:greedy} 
\end{figure}

\begin{table*}
\caption{\label{tab:phicases}Parameters for different injected values of $\phi$ for the analysis described in Sec.~\ref{sec:retrograde}. The standard deviation of the Gaussian noise $\sigma$ is set such that the corresponding post-peak SNR is fixed at $\rho_{\rm RD} = 10$. The time of the peak strain $t_{\rm peak}$ is quoted relative to $t_0$ [Eq.~(\ref{eq:t0})]. The truncation time $t_{\rm cut}$ is set to be the time when the error between the best template and the injection waveform $|\hat{h}^{RD} - h^{\rm N}|$ falls below $\sigma / 10$. $\rho_{\rm cut}$ is the remaining SNR in the signal after truncating the data. }
\begin{ruledtabular}
\begin{tabular}{c|cccc|cccc|cccc|cccc|cccc}
 &\multicolumn{4}{c}{$a/M = 0.1$} & \multicolumn{4}{c}{$a/M = 0.3$} & \multicolumn{4}{c}{$a/M = 0.5$} & \multicolumn{4}{c}{$a/M = 0.7$} & \multicolumn{4}{c}{$a/M = 0.9$} \\
 $\phi$ & $\sigma$ & $t_{\rm peak}$ & $t_{\rm cut}$ & $\rho_{\rm cut}$  & $\sigma$ & $t_{\rm peak}$ & $t_{\rm cut}$ & $\rho_{\rm cut}$ & $\sigma$ & $t_{\rm peak}$ & $t_{\rm cut}$ & $\rho_{\rm cut}$ & $\sigma$ & $t_{\rm peak}$ & $t_{\rm cut}$ & $\rho_{\rm cut}$ & $\sigma$ & $t_{\rm peak}$ & $t_{\rm cut}$ & $\rho_{\rm cut}$\\ \hline
 0 & 3.2 & -3.4 & 8 & 3.5 & 2.8 & -3.5 & 8 & 4.4 & 3.6 & -12.4 & 5 & 3.5 & 3.0 & -10.0 & 4 & 3.3 & 2.4 & -9.4 & 17 & 2.1 \\
 $\pi/3$ & 4.1 & -11.5 & 11 & 3.0 & 3.8 & -11.5 & 6 & 4.0 & 3.1 & -7.9 & 7 & 3.2 & 2.5 & -6.0 & 1 & 6.4 & 3.4 & -17.7 & 15 & 1.9 \\
 $2\pi/3$ & 3.7 & -7.4 & 8 & 4.2 & 3.3 & -7.4 & 8 & 3.5 & 2.7 & -4.2 & 3 & 6.3 & 3.6 & -14.8 & 5 & 3.0 & 2.9 & -13.3 & 18 & 1.2 
\end{tabular}
\end{ruledtabular}
\end{table*}

In Table~\ref{tab:greedy}, we show all models in the final set $S$ at the termination of the algorithm. {At $\rho_{\rm RD} = 25$, we find that largest signal-to-noise log Bayes factor (the evidence normalized by the Gaussian noise evidence) for the model $M=\lbrace (2,-1),(2,-2) \rbrace$, with $\log B_{\rm max} = 59.0$. We find the smallest evidence for the model $M = \lbrace (3,-3) \rbrace$, with $\log B_{\rm min} = 37.8$. While $N = 2$ modes are generally supported by the data, all $N \geq 3$ models are excluded from the final set of models. Since there is insufficient evidence for adding a third mode, our algorithm terminates at this point. We verify that the Bayes factor decreases with each added mode that is not supported by the data. Continuing the greedy algorithm, the maximum evidence across all $N = 3,4,5,6$ models decreases as $52.5,46.6,43.4,41.4$, respectively.}

In Fig.~\ref{fig:greedy}, we show the posterior distributions for the fitting amplitudes $F_{2-1}$ and $F_{2-2}$. We plot the best $N = 2$ model as ranked by the Bayesian evidence, $M=\lbrace (2,-1),(2,-2) \rbrace$, as well as the $N = 6$ model, $M = \lbrace (2,2),(2,1),(2,0),(2,-1),(2,-2),(3,-3) \rbrace$. This example highlights the difficulty in recovering the true mode content without source-informed priors. The $N=2$ models recover the mode amplitudes with significant bias, and the $N=6$ model recovers the mode amplitudes consistently but at the expense of much wider credible intervals. {For the $N=6$ model, the median and $2\sigma$ credible interval of the marginal posteriors are $F_{2-1} = -270.0^{+719.2}_{-451.6}$ and $F_{2-2}=179.5^{+148.6}_{-239.0}$.} Furthermore, the $N \geq 3$ models do not significantly improve the fit, according to the Bayesian evidence. In comparison, constraints on the fitting amplitudes are significantly improved by using source information. To generate posterior samples in $F_{\ell m}$, we post-process the posterior samples in $\lbrace I, \delta, d_L, \iota, \phi \rbrace$ from Sec.~\ref{sec:casestudy} and map them to the fitting amplitudes using Eq.~(\ref{eq:fitpars}).

\subsection{\label{sec:retrograde} Test for retrograde motion}

At lower SNR $\rho_{\rm RD} = 10$, the individual source parameters have broad credible intervals and cannot be constrained as well as demonstrated in Fig.~\ref{fig:thinc_delta}. In this regime, we examine how the spin of the central black hole impacts the measurement of the mode amplitudes and the plunge geometry.

The parameters we use to describe the plunge geometry are defined with respect to the spin axis of the black hole. As the spin of the black hole decreases, the spacetime becomes spherically symmetric, and the plunge parameters $I$ and $\delta$ become degenerate with the extrinsic parameters $\iota$ and $\phi$. When the spin is sufficiently small, the ringdown spectrum as excited by a ``misaligned" plunge will be degenerate with that of a spin-aligned $(I = 0)$ plunge. 

For a given plunge geometry, we can estimate the spin value required to break this degeneracy and measure the plunge parameters. As an example, consider the ringdown waveform from a retrograde plunge for a fixed geometry $I$ and $\delta$. The degree of degeneracy can be characterized by the preference for a model which assumes the system came from a retrograde plunge $\pi/2 < I \leq \pi$ over a model which assumes the system came from a prograde plunge $0 \leq I < \pi/2$. This can be calculated with the Bayes factor, 
\begin{eqnarray} \label{eq:bayesfactor}
    \log \mathcal{B}^{\rm retrograde}_{\rm prograde} = \log \left (\mathcal{Z}_{\rm retrograde} / \mathcal{Z}_{\rm prograde} \right),
\end{eqnarray}
where $\mathcal{Z}_{\rm retrograde}$ ($\mathcal{Z}_{\rm prograde}$) is the total evidence of a model over a prior space with $I$ restricted to only retrograde (prograde) orientations.  If $\log B^{\rm retrograde}_{\rm prograde} \approx 0$, both plunge orientations are equally supported by the data. Alternately, if $\log B^{\rm retrograde}_{\rm prograde} \gtrsim 5$, the retrograde plunge model is strongly preferred by the data. As the spin of the black hole increases, the degeneracy breaks and the Bayes factor should increase.

We inject a waveform from a retrograde equatorial plunge $I = \pi$ with a face-on orientation $\iota = 0$. The other parameters, except for $a$ and $\phi$, are set to the same values as in Secs.~\ref{sec:casestudy} and \ref{sec:Nmode}. The dominant ringdown mode is the $(2,-2,0)$, with subdominant contributions from $(3,-2,0)$ and $(4,-2,0)$ modes. In this case, we can interpret the Bayes factor in a rather simple way --- the degree to which the injected retrograde $(2,-2)$ mode can be emulated by prograde modes which are not actually in the data. We run a parameter estimation for each of the black hole spins we have available $a/M = 0.1, 0.3, 0.5, 0.7, 0.9$.

The viewing angle $\phi$ changes the interference between the ringdown modes. This can have a large impact on the parameter estimation when the data only contain a few measurable cycles. In order to marginalize over this effect, we calculate the average Bayes factor over several different injected $\phi$ values. We use several different injection values $\phi_n = n \pi / 3$, with $n = 0,1,2$. Since only the $m = -2$ modes are observable (our injection waveform is face-on $\iota = 0$), we only need to sample in the range $0 \leq \phi \leq \pi$, as the phase of each mode depends on $e^{\pm i m \phi}$. 

For each $\phi$ value, we set the SNR of the injected waveform to $\rho_{\rm RD} = 10$ by adjusting the noise level $\sigma$. We then calculate the analysis start time $t_{\rm cut}$, and post-cut SNR $\rho_{\rm cut}$, using the same prescription as in Sec.~\ref{sec:casestudy} (when the residual between the best template and the injected waveform drops below $\sigma / 10$).

In Table.~\ref{tab:phicases}, we summarize the waveform and analysis parameters that change with $\phi$. For spins up to $a/M \leq 0.7$, the post-cut SNR can vary between $3 \leq \rho_{\rm cut} \leq 6.4$. At $a/M = 0.9$, the post-cut SNR is smaller at $1.2 \leq \rho_{\rm cut} \leq 2.1$. This is because our ringdown model omits overtones, which are more prominent at high spins. This means that for $a/M = 0.9$, the residual $|\hat{h}^{RD} - h^{\rm N}|$ is dominated by overtones for much longer than the other cases we examine, and does not fall below the $\sigma / 10$ level until much of the signal has already decayed.

\begin{figure}[t]
    \centering
    \includegraphics[width=.4\textwidth]{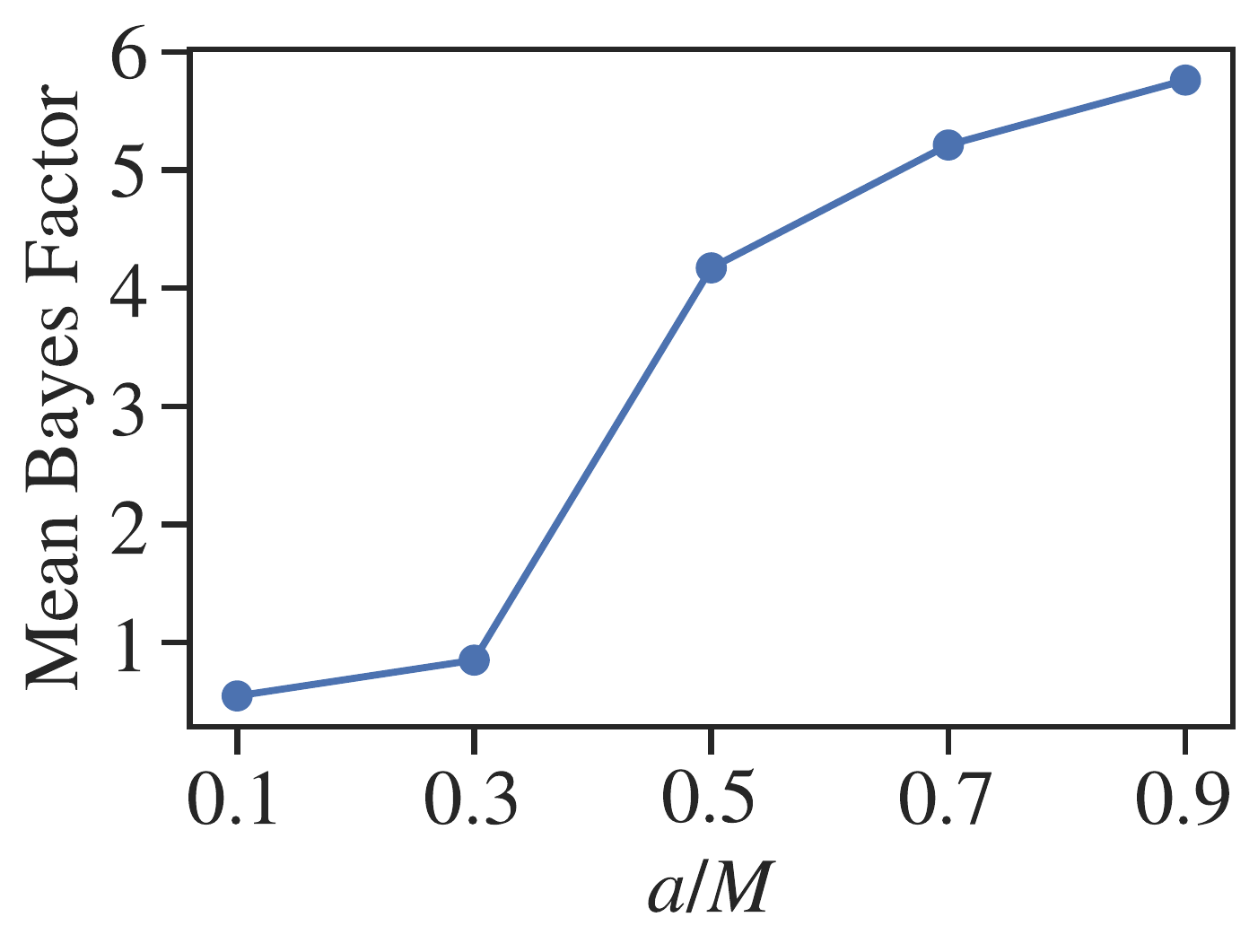}
    \caption{The log Bayes factor between retrograde and prograde plunge models at different black hole spins. We inject the waveform from a face-on, retrograde equatorial plunge ($\iota = 0$, $I = \pi$) with post-peak SNR $\rho_{\rm RD} = 10$. The model evidences $ \mathcal{Z}_{\rm retrograde}$ and $\mathcal{Z}_{\rm prograde}$ at each spin are averaged over the three analyses with different injected $\phi$ values. Since different injected $\phi$ values lead to different interference between excited modes and different $t_{\rm peak}$ times, in order to keep $\rho_{\rm RD} = 10$ fixed we must compensate by adjusting the noise and fitting windows, as shown in Table.~\ref{tab:phicases}.}
    \label{fig:bayes} 
\end{figure}

The average Bayes factor comparing retrograde and prograde priors is plotted in Fig.~\ref{fig:bayes}. To calculate the average Bayes factor, we first run the parameter estimation for the prograde and retrograde models to find the Bayes factor for each $\phi_n$ case, then take the average overall all cases. As expected, the ability to distinguish the retrograde modes increases with spin. At the highest spin case we consider $a/M = 0.9$ the SNR of truncated data is the lowest at $\rho_{\rm cut} \approx 2$. Even at this lower SNR, the retrograde modes are distinct from the prograde modes; the Bayes factor indicates that the prograde plunge models are about 400 times less probable than the retrograde models.  

\section{\label{sec:Discussion} Discussion}

In this work, we describe a ringdown waveform model consisting of QNMs for generically misaligned precessing black holes in the large mass ratio limit. We use this waveform model to demonstrate a proof-of-principle measurement of the QNM amplitudes and plunge geometry by conducting a Bayesian inference with white Gaussian noise. This work motivates future efforts to further characterize and map the QNM excitation from compact binary coalescences. 

While aligned or anti-aligned binary mergers primarily excite either the prograde or retrograde modes, highly misaligned sources tend to excite both. In order to model the greater variety, generic ringdown models are needed that do not rely on a standard set of modes. The payoff for such efforts will come when we have detectors that can measure misaligned coalescences and sources with large enough mass ratio that the final spin is not dominated by orbital angular momentum at plunge. 

In this study, we also demonstrate the benefit of incorporating source information to constrain the mode amplitudes and phases in terms of the source parameters. Simply adding additional, unconstrained QNMs to a model will lead to suboptimal measurements. To illustrate this, we analyze the ringdown from a plunge with a spin of $a/M = 0.5$ and spin-orbit misalignment of $137^\circ$ that excites all $(2,m)$ modes. Without a source model for the mode excitation, the number and variety of modes to include in the ringdown model must be determined. We make this determination through Bayesian model selection and find maximal evidence for a $N = 2$ mode model, but also find the same model leads to biased measurements of the mode amplitudes. On the other hand, an $N = 6$ mode model leads to unbiased measurements but at the expense of wide and uninformative credible intervals.

The parameters we use to describe the plunge geometry are spherical coordinates defined with respect to the spin axis of the larger black hole. As the spin decreases, this parameterization becomes degenerate with the extrinsic emission angles, and the distinction between the retrograde and prograde modes vanish. While fully generic ringdown models may be needed for larger spins and SNRs, these results suggest that misaligned systems with sufficiently small remnant spin may still be effectively modeled using only an aligned-spin ringdown model, plus some extrinsic rotation. 

Although full inspiral-merger-waveform waveforms models already exist that effectively model the post-peak ringdown, one benefit of mapping out the mode excitation is the prospect for conducting tests of GR based on the mode amplitudes. In this work, we used such a mapping to directly sample in source parameter space ($I$ and $\delta$), but another possibility is to instead construct amplitude priors and sample in amplitude space. Sampling in amplitude space would allow for the flexibility to relax our prior and perform consistency checks between direct measurements of the mode amplitudes with GR predictions based on initial binary parameters. The detection of significant deviations from these expected amplitudes or phases, possibly in conjunction with deviations in the mode frequencies, may indicate physics beyond GR. 

\section*{Acknowledgments}

Our work on this problem was supported at MIT by NSF Grant No.~PHY-2110384. We would like to thank Lionel London, Maximiliano Isi, and Salvatore Vitale for helpful discussions on this work. G.K. acknowledges support from NSF Grants No.~PHY-2106755 and No.~DMS-1912716 and feedback from Tousif Islam and Scott Field. Simulations were performed on the MIT Lincoln Labs {\em SuperCloud} GPU supercomputer supported by the Massachusetts Green High Performance Computing Center (MGHPCC) and ORNL SUMMIT under allocation AST166. Corner plots were generated using \texttt{corner} package \cite{Foreman-Mackey2016Corner.py:Python}.

\appendix

\section{Controlling modeling errors in mode extraction}
\label{app:errors}

In this Section, we discuss how we improve the accuracy of fits in the presence of unmodeled contributions. When the data are well described by a known set of QNMs, the recovered mode amplitudes, {after shifting the amplitudes to their values at the fiducial time according to Eq.~(\ref{eq:timeshift})}, should not be sensitive to when those QNMs are fit provided the data sufficiently constrain those modes. Most of the waveforms in our dataset demonstrate this consistency, but there are some cases where we find the QNM fits are inconsistent, varying over fitting time.  

\begin{figure}[ht]
    \centering
    \includegraphics[width=.45\textwidth]{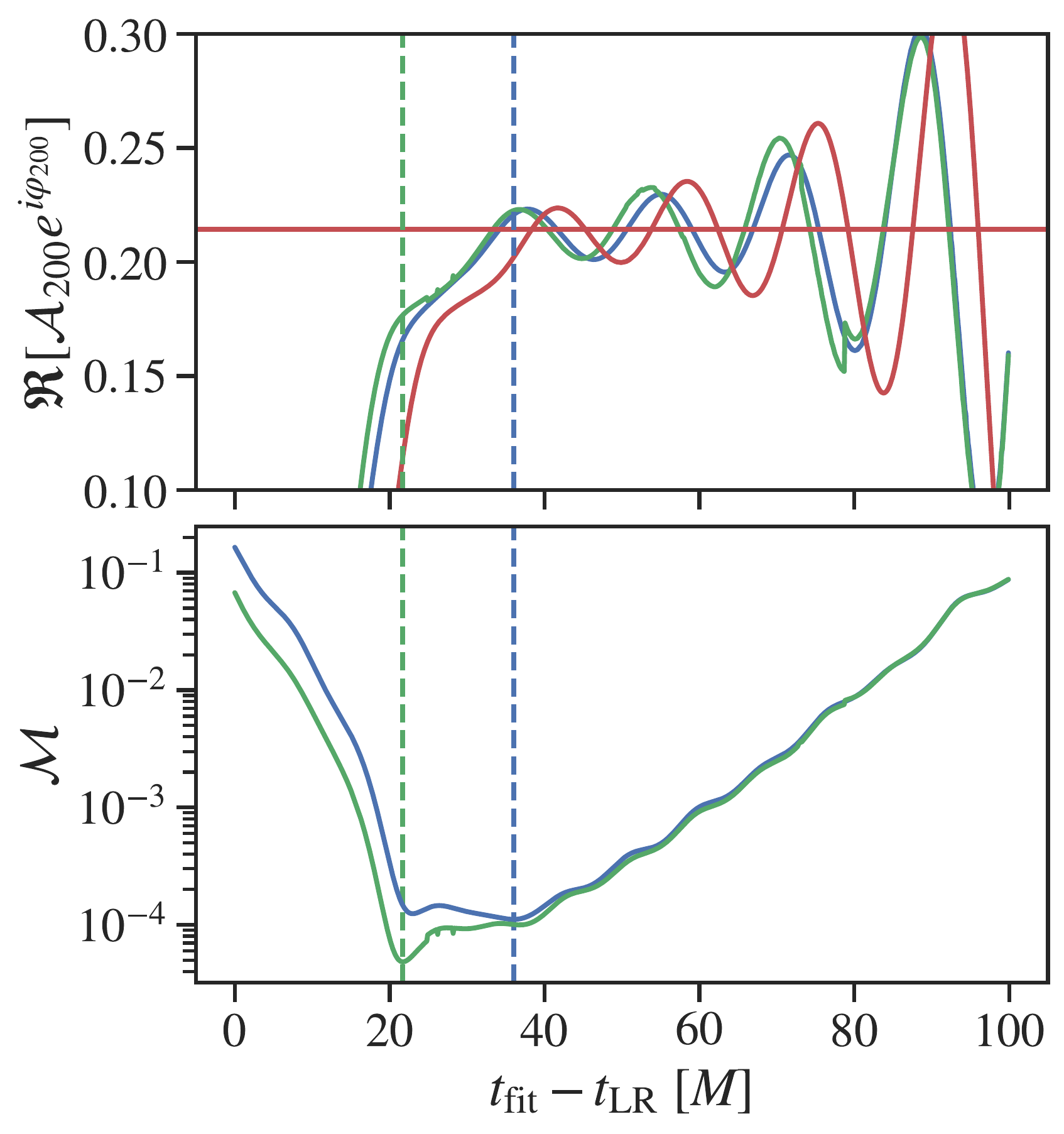}
    \caption{The $(2,0,0)$ mode amplitudes extracted from the $h^{\rm N}_{00}$ spherical mode in the presence of suspected numerical errors. {We plot the fit time relative to the fiducial time $t_0 = t_{\rm LR}$ as defined in Eq.~(\ref{eq:t0}).} In the top panel, we plot the fit results from a least squares algorithm as a function of fitting window $\lbrack t_{\rm fit}, 150M \rbrack $ with a model containing $N = 1$ QNM pairs (blue line) and $N = 3$ QNM pairs (green line). We also plot the fit results from the LKAH algorithm with $N = 3$ QNM pairs (red line) as a function of the instantaneous fit time $t_{\rm fit}$. The horizontal line represents the average of the fit results from the LKAH algorithm over the time window which minimizes the standard deviation of the instantaneous mode amplitudes, as described in Ref.~\cite{Lim2019}. In the bottom panel, we show the mismatch of the two least squares fits, where the vertical lines indicates the fitting windows at which the mismatches are minimized.}
    \label{fig:mismatch}
\end{figure}

In a least-squares fit, the ``fitting time" is equivalent to the choice of fitting window. For a spherical-to-spheroidal basis transformation as performed in Ref.~\cite{Lim2019} (which we refer to as the LKAH algorithm), ``fits" may be performed at each point in time. In both procedures, the fitting time can be considered a hyperparmeter. As discussed in Sec.~\ref{ssec:extraction}, we advocate for taking an average of fits conducted across a range of fitting times. Such an average helps if there are contributions to the waveform that are not QNMs (e.g. numerical errors, unmodeled modes), and if there is evidence of overfitting to these contributions leading to inconsistent mode amplitudes.

\begin{figure}[t]
    \centering
    \includegraphics[width=.45\textwidth]{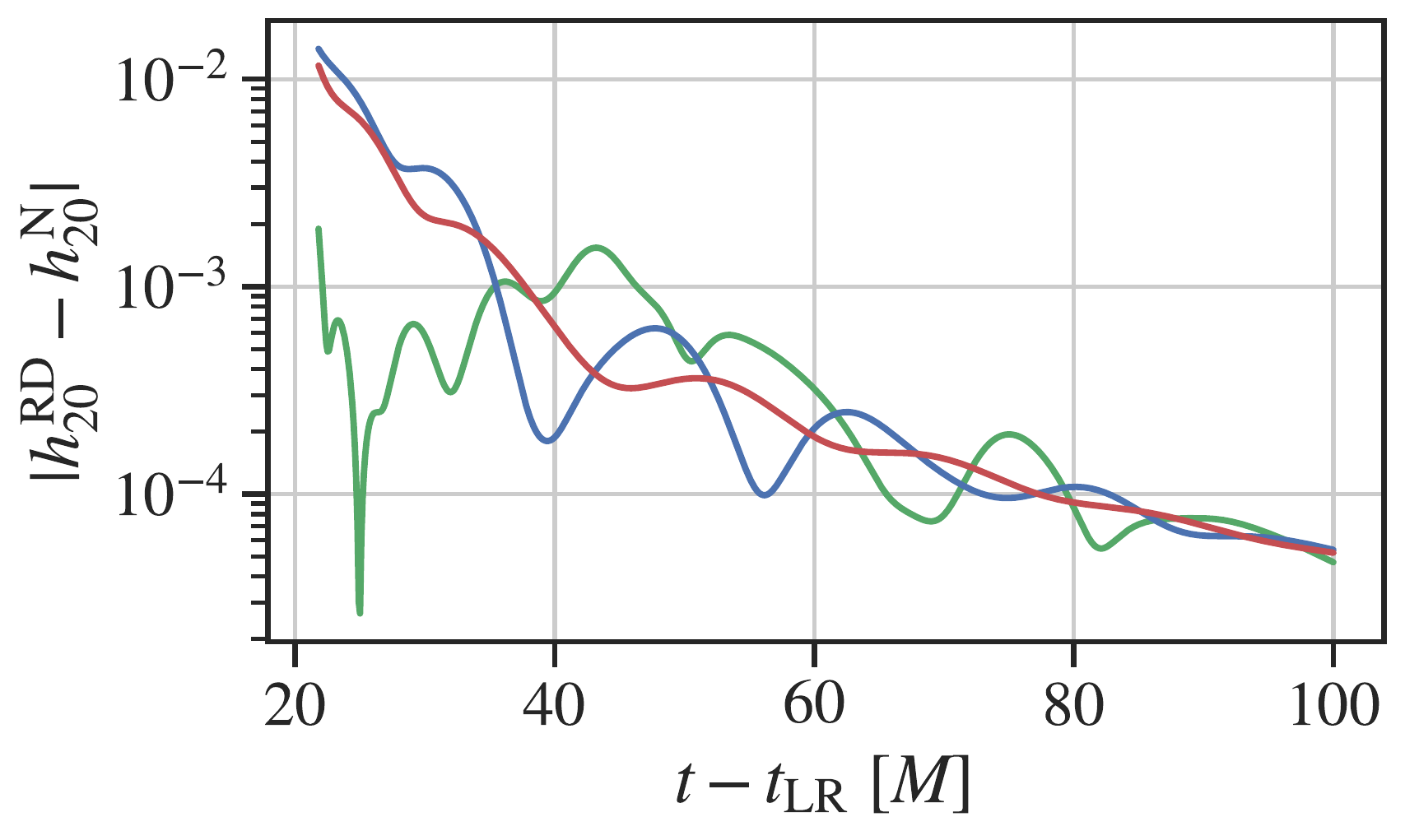}
    \caption{The magnitude of the fit residuals between perturbation theory models and the spherical mode data. For the least squares fits, we plot the $h_{20}^{{\rm RD},1}$ and $h_{20}^{{\rm RD},3}$ models with fit amplitudes taken from the lowest mismatch fit Fig.~\ref{fig:mismatch}. For the LKAH fit, we plot the $h_{20}^{{\rm RD},3}$ model with the fit amplitudes taken from the average over fit times.}
    \label{fig:errors}
\end{figure}

We now demonstrate this procedure on a waveform where we find the  variations in the fits are especially large: the $(\ell,m) = (2,0)$ multipole for the $a/M = 0.1$, $I = 25^\circ$, $\theta_{\rm fin} = 87.8$, $\dot{\theta}_{\rm fin} < 0$ plunge. Consider a least-squares algorithm which minimizes the squared residual of the $(\ell,m) = (2,0)$ spherical mode over a time window,
\begin{equation}
    \epsilon = \sum\limits_{t_i = t_{\rm fit}}^{150\ M} \left |h^{\rm N}_{20}(t_i) - h^{{\rm RD},N}_{20}(t_i) \right |^2.
\end{equation}    
$h^{\rm N}_{20}$ is the $(2,0)$ spherical mode, the output of our time-domain Teukolsky solver. We fit the data with two different models with $N=1$ and $N=3$ QNM pairs, respectively, where
\begin{eqnarray}
    & h^{{\rm RD},N}_{20} & =  \\
    & & \sum_{k=2}^{N+1}  \left[ a_{02k0}(t) \mathcal{A}_{k00} e^{i \varphi_{k00}} + a_{02k0}'(t) \mathcal{A}'_{k00} e^{i \varphi'_{k00}} \right].  \qquad   
\end{eqnarray}
The coefficients $a_{02k0}(t)$ and $a_{02k0}'(t)$ contain the spherical-spheroidal overlap coefficients and the time dependence of each QNM. They are defined in Eq.~(3.8) in Ref.~\cite{Lim2019}.

\begin{figure*}[ht]
    \centering
    \includegraphics[width=.9\textwidth]{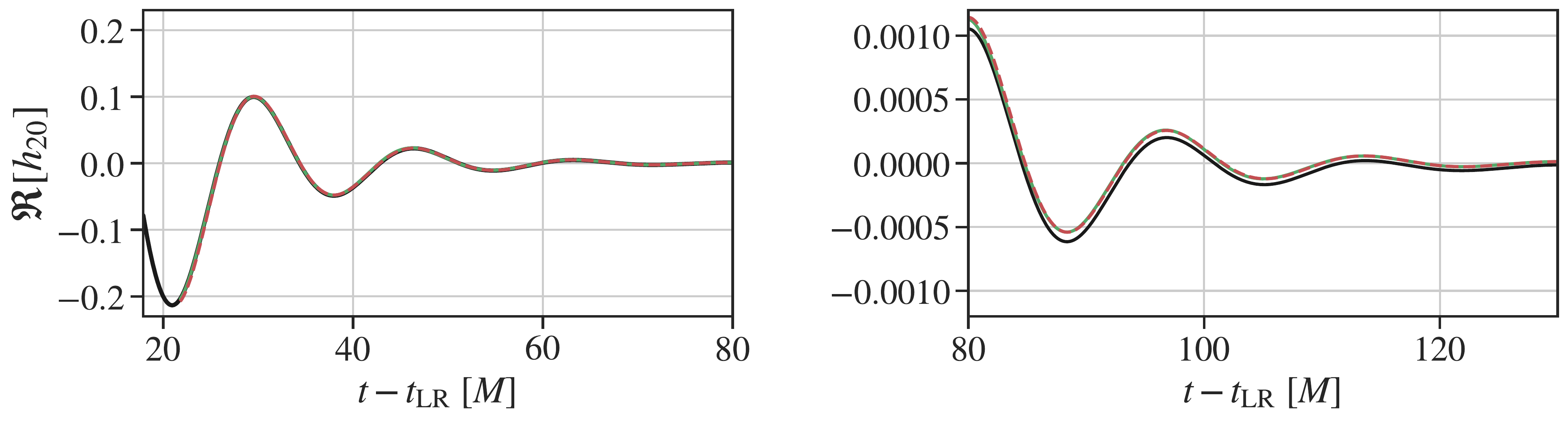}
    \caption{The spherical mode data (black) and the perturbation theory models $h_{00}^{{\rm RD},3}$ (colors are the same as in Figs.~\ref{fig:mismatch} and \ref{fig:errors}). In the late ringdown, the two perturbation theory models agree with each other but do not agree with the data at the $10^{-5}$ level. }
    \label{fig:noise}
\end{figure*}

In the top panel of Fig.~\ref{fig:mismatch}, we show a sequence of fitting results from a least-squares fit performed over successive fitting start times $t_{\rm fit}$, for the $N=1$ mode model and the $N = 3$ mode model. For comparison, we show the results from an additional non-least-squares fit, which also uses the same $N = 3$ pairs of QNMs but fitted using the LKAH algorithm (where the fitting time represents a local point and not a window). For all models, there are oscillations in the mode amplitudes as the fitting time is changed. We expect agreement between the $N = 1$ and $N = 3$ models because the higher order QNMs represent very small corrections; not only is the excitation of the higher order QNMs intrinsically weaker but the overlap between the $(k,m)=(3,0)$ and $(k,m)=(4,0)$ spheroidal harmonics with the $(\ell,m) = (2,0)$ spherical harmonic is small:
\begin{equation}
    |\mu_{0230}| = 1.5\times10^{-2}, \qquad |\mu_{0240}| =  8\times10^{-5},
\end{equation}
where we define the overlap $\mu_{m\ell k n}$ in Eq.~(3.5) in LKAH.

In the bottom panel of Fig.~\ref{fig:mismatch}, we plot the the mismatch for the $N=1$ and the $N=3$ fits across different fitting times. The key detail is that by simply using the fits which minimize the mismatch, the $N=1$ and $N=3$ fits will be inconsistent with each other by about 20\%, which contradicts our expectation that the models should agree.

We suspect that the minimum possible mismatch is lower for the $N=3$ model mainly due to overfitting at early times, rather than being more descriptive of the underlying mode content. In Fig.~\ref{fig:errors}, we plot the residual between the data and the $N=3$ and $N = 1$ fits with the lowest mismatches, respectively. We also plot the residual between the data and the fit performed with the LKAH algorithm. The residual for the $N=3$ (least squares fit) is the lowest at early times, but then increases substantially between $25 \leq t/M \leq 50$ and exhibits the largest error across all models between $40 \leq t/M \leq 60$. If the $N=3$ model were truly a better description of the data, we would expect the agreement between the model and the data to improve at later times. Instead, we see behavior consistent with overfitting. 

As seen in Fig.~\ref{fig:mismatch}, the fitted mode amplitudes oscillate about a constant after times $t/M \gtrsim 40 $. For all three models, the same constant can be recovered by taking the hyperparameter average, which resolves the inconsistency. Interestingly, the amplitudes of these oscillations, as well as the mismatch, increase with fitting time after $t/M \gtrsim 40$. We identify this behavior with low-frequency numerical errors that cannot be represented by QNMs. We show the waveform in Fig.~\ref{fig:noise}. At times $t/M > 80$, the $N=3$ least squares model and the LKAH fit agree with each other, but neither model precisely matches the data at the $10^{-5}$ level. The residuals appear one-sided over this time interval, and thus cannot be modeled with a sum of sinusoidal (zero-average) functions. Nevertheless, whenever we observe this behavior in our data, we are still able to perform a hyperparameter average which leads to consistent fits across different fitting procedures and fitted modes.

\bibliography{ref} 
\end{document}